\documentclass[fleqn,usenatbib]{mnras}

\usepackage[T1]{fontenc}

\DeclareRobustCommand{\VAN}[3]{#2}
\let\VANthebibliography\thebibliography
\def\thebibliography{\DeclareRobustCommand{\VAN}[3]{##3}\VANthebibliography}

\usepackage{graphicx}	
\usepackage{amsmath}	
\usepackage{amssymb}	
\usepackage{diagbox} 
\usepackage{caption}
\usepackage{subcaption}
\usepackage{tablefootnote}
\usepackage{newtxtext,newtxmath}

\title[PSR~J1910$-$5320]{Mass estimates from optical modelling of the new TRAPUM redback PSR~J1910$-$5320}
\author[O. G. Dodge et al.]{O. G. Dodge,$^{1}$\thanks{E-mail: oliver.dodge@manchester.ac.uk}
R. P. Breton,$^{1}$
C. J. Clark,$^{2,3}$
M. Burgay,$^{4}$
J. Strader,$^{5}$
K.-Y. Au,$^{6}$
E.~D.~Barr,$^{7}$
S. Buchner,$^{8}$\newauthor 
V. S. Dhillon,$^{9,10}$
E. C. Ferrara,$^{11,12,13}$
P. C. C. Freire,$^{7}$
J.-M. Grie{ss}meier,$^{14,15}$
M. R. Kennedy,$^{16,1}$\newauthor
M. Kramer,$^{7,1}$
K.-L. Li,$^{6}$ 
P. V. Padmanabh,$^{2,3,7}$
A. Phosrisom,$^{1}$
B. W. Stappers,$^{1}$
S. J. Swihart,$^{17}$\newauthor
T. Thongmeearkom$^{1,18}$
\\
$^{1}$Jodrell Bank Centre for Astrophysics, The University of Manchester, Alan Turing Building, Manchester, M13 9PL, UK\\
$^{2}$Max Planck Institute for Gravitational Physics (Albert Einstein Institute), D-30167 Hannover, Germany\\
$^{3}$Leibniz Universität Hannover, D-30167 Hannover, Germany\\
$^{4}$INAF - Osservatorio Astronomico di Cagliari, via della Scienza 5, 09047 Selargius (CA), Italy\\
$^{5}$Center for Data Intensive and Time Domain Astronomy, Department of Physics and Astronomy, Michigan State University, East Lansing, MI 48824, USA\\
$^{6}$Department of Physics, National Cheng Kung University, No. 1 University Road, Tainan City 70101, Taiwan\\ 
$^{7}$Max-Planck-Institut f\"{u}r Radioastronomie, 53121 Bonn, Germany\\
$^{8}$South African Radio Astronomy Observatory, 2 Fir Street,Cape Town, 7925, South Africa\\
$^{9}$Department of Physics and Astronomy, University of Sheffield, Sheffield S3 7RH, UK \\
$^{10}$Instituto de Astrof\'{i}sica de Canarias, E-38205 La Laguna, Tenerife, Spain \\
$^{11}$Department of Astronomy, University of Maryland, College Park, MD, 20742, USA\\
$^{12}$Center for Research and Exploration in Space Science \& Technology II (CRESST II), NASA/GSFC, Greenbelt, MD 20771, USA\\
$^{13}$NASA Goddard Space Flight Center, Greenbelt, MD 20771, USA\\
$^{14}$LPC2E - Universit\'{e} d'Orl\'{e}ans /  CNRS, 45071 Orl\'{e}ans cedex 2, France\\ 
$^{15}$Observatoire Radioastronomique de Nan\c{c}ay (ORN), Observatoire de Paris, Universit\'{e} PSL, Univ Orl\'{e}ans, CNRS, 18330 Nan\c{c}ay, France\\
$^{16}$School of Physics, Kane Building, University College Cork, Cork, Ireland \\
$^{17}$Institute for Defense Analyses, 730 E Glebe Rd, Alexandria, VA 22305, USA\\
$^{18}$ National Astronomical Research Institute of Thailand, Don Kaeo, Mae Rim, Chiang Mai 50180, Thailand\\
}
\date{Accepted XXX. Received YYY; in original form May 2021}
\pubyear{2023}

\begin{document}
\label{firstpage}
\pagerange{\pageref{firstpage}--\pageref{lastpage}}
\maketitle

\begin{abstract}
Spider pulsars continue to provide promising candidates for neutron star mass measurements. Here we present the discovery of PSR~J1910$-$5320, a new millisecond pulsar discovered in a MeerKAT observation of an unidentified \textit{Fermi}-LAT gamma-ray source. This pulsar is coincident with a recently identified candidate redback binary, independently discovered through its periodic optical flux and radial velocity. New multi-color optical light curves obtained with ULTRACAM/NTT in combination with MeerKAT timing and updated SOAR/Goodman spectroscopic radial velocity measurements allow a mass constraint for PSR~J1910$-$5320. \texttt{Icarus} optical light curve modelling, with streamlined radial velocity fitting, constrains the orbital inclination and companion velocity, unlocking the binary mass function given the precise radio ephemeris. Our modelling aims to unite the photometric and spectroscopic measurements available by fitting each simultaneously to the same underlying physical model, ensuring self-consistency. This targets centre-of-light radial velocity corrections necessitated by the irradiation endemic to spider systems. Depending on the gravity darkening prescription used, we find a moderate neutron star mass of either $1.6\pm0.2$ or $1.4\pm0.2$ $M_\odot$. The companion mass of either $0.45\pm0.04$ or $0.43^{+0.04}_{-0.03}$ $M_\odot$ also further confirms PSR~J1910$-$5320 as an irradiated redback spider pulsar.
\end{abstract}

\section{Introduction}
\label{Intro}

The fastest subset of pulsars are known as millisecond pulsars (MSPs), quite simply due to their millisecond spin periods. In addition to their blistering rotations, MSP periods also decay slowly relative to other pulsars due to surface magnetic fields several orders of magnitudes lower than the general pulsar population. Their extreme characteristics are thought to be attained in a suitably exotic manner; the recycling scenario ascribes the `spin-up' of an old, slow neutron star to the accretion of mass from a binary companion. This transfers angular momentum onto the neutron star, accelerating its spin speed. Given a suitably long period of mass transfer, the neutron star may be spun up to millisecond periods \citep{Alpar1982, Bhattacharya1991}. 

Given the recycling scenario, spinning up an MSP requires a companion. However, since around 20\% of known MSPs are isolated \citep{Jiang2020}, one needs to explore how these seemingly  lost their companion. The discovery of the first `black widow' MSP by \citet{Fruchter1988} presented one possible formation mechanism, and established the `spider' pulsar sub-class of MSPs. Typically a spider system pairs a low-mass, non-degenerate companion with an MSP in a compact ($P_{B}$ < 24 hours) orbit. The companion is tidally locked to the pulsar, thus the irradiating pulsar wind heats one face whilst the opposite side remains cooler \citep{D&S1998}. This irradiation ablates material from the companion which often results in eclipsing of the pulsar's beam at radio frequencies \citep[see, e.g.,][]{Polzin2020}, as well as leading to their nicknames - associating their cannibalistic tendencies with arachnid analogues. Though spiders initially appeared a promising route to isolated MSPs, it still remain highly uncertain as to whether full evaporation within a Hubble time is a realistic option \citep[see, e.g.,][]{Stappers1996,Polzin2020,Kandel2021}. In any case, they provide fascinating environments to study the pulsar wind and high energy particle physics.

Spider pulsars are typically split into two a categories based on their companion mass: black widows with extremely low mass ($M_C < 0.05 M_\odot$) and redbacks with higher companion masses ($M_C \gtrsim 0.1 M_\odot$) \citep{Roberts2013}. Black widows normally have single peaked light curves over an orbital period, as the impinging irradiation flux dominates the companion star's base temperature. Redbacks light curves can also often exhibit strong irradiaton, though unlike black widows it is not ever-present as their base temperatures are higher. Thus, the relative contribution to their light curves of ellipsoidal modulation caused by the tidal distortion of the star is important and produces two peaks per orbital period \citep[see][for discussion on the interplay between irradiation and tidal effects in redbacks]{Turchetta2023}. Three redbacks, known as transitional millisecond pulsars (tMSPs), were witnessed to switch between MSP (radio-loud) and accreting low-mas X-ray binary (LMXB) states, with each state typically lasting a few years or more. tMSPs are hailed as providing clear evidence for the recycling scenario described above \citep{Archibald2009, Papitto2014, Bassa2014, Stappers2014}. 

Constraining the neutron star equation of state (EoS), through neutron star mass measurements \citep{Ozel&Freire2016}, fuels a great deal of interest in spider pulsars. \citet{Linares2019} has demonstrated that spiders often host particularly massive neutron stars, with several contending to be the most massive neutron star observed. The original black widow, PSR B1957+20 for a time seemed the heaviest known neutron star, clocking in at 2.4 M$_\odot$ \citep{VK2011}. Improved knowledge and data around $\gamma$-ray eclipsing in spiders has since revised this measurement down significantly \citep{Clark2023b}, but the promise of massive neutron stars in spider systems remains. There are many EoS model contenders, each predicting a maximum possible neutron star mass. Thus by observing and measuring massive neutron stars, any EoS predicting a maximum mass below that of the most massive known neutron star can be discarded. The binary nature of spiders where both components can be studied separately therefore provides a convenient avenue to constraining neutron star masses. Radio timing provides the orbital period and pulsar radial velocity, while optical observations can determine inclination and companion radial velocity from photometric and spectroscopic modelling, respectively. Once put together, these can constrain the masses in the system. This then motivates the work in this paper: any new spider to be characterised provides valuable mass measurements and a potential to constrain the EoS. Whilst there are a number of systematics and assumptions inherent to optical modelling when compared with other neutron star mass measurements \citep[see][]{Ozel&Freire2016}, \citet{Romani2021}, \citet{Kennedy2022} and \citet{Clark2021} clearly demonstrate the potential spiders have for precise mass determinations.

Spectroscopic modelling of spiders is relatively novel field, certainly when compared with its photometric counterpart. Both sides of spider modelling are far from complete providing complete descriptions of the companion, with spectroscopic modelling in particular suffering from its extreme computational expense. Aside from technical concerns, the fundamental complications when measuring the radial velocity in spider binaries from observations are summarised as \textit{``centre-of-light''} effects. Determining the binary mass ratio, requires to combine the well-measured pulsar's projected semi-major axis with a value of the companion's projected centre-of-mass radial velocity. However the radial velocities derived from observed spectroscopy track the centre of light of the particular line or set of lines observed. Indeed, the non-uniform temperature and non-spherical shape of the companion imply that the strength of a line may vary greatly across its surface, which translates into a line velocity that is offset from the center of mass, therefore producing a different projected radial velocity amplitude but also an orbital profile which may depart slightly from the perfect  function expected from a circular orbit.

Several approaches have been used to connect the observed radial velocities to the correct centre-of-mass radial velocity amplitude. \citet{VK2011} and \citet{Romani2021} both produced synthetic radial velocity curves which are then fitted to the observed curve to estimate the correction factor. \citet{Linares2018}, on the other hand, takes a more empirical approach in which observed line species are assessed to originate from the hotter dayside or colder nightside of the companion based on the temperature at which they are produced. In this way, they can `bracket' line velocities to lie between the true centre-of-mass and the maximal extent of the star in either direction. Finally, \citet{Kennedy2022} implemented the ultimate step in producing full synthetic spectra which are directly fitted to the raw observed spectroscopy. This modelling of the photometry and spectroscopy ensures the necessary centre-of-light corrections are intrinsically embedded in the line profile which is self-consistent with the heating model at any given parameters.

\par Follow up observations are fruitful in various wavelengths; \citet{Ray2013} reported the discovery of 43 new MSPs, many of which were spiders, from the first generation of deep radio searches targeting unassociated \textit{Fermi}-LAT sources. The population has kept growing since, with the latest \textit{Fermi}-LAT survey reporting at least 110 MSPs discovered in this fashion \citep{Fermi3}. In addition to these, \citet{TRAPUMSurv} detailed a new MeerKAT L-band survey of LAT sources in which 9 new MSPs were found among 79 \textit{Fermi}-LAT sources, including two new redbacks. Optical searching of similar fields, with or without prior radio search, can also produce new spider candidates by looking for the signature orbital modulation in the light curves described earlier, with spectroscopy possibly providing further evidence through the system's mass function \citep[see, e.g.,][]{Strader2015, Strader2016, Swihart2022}.

One such recent discovery is that of a candidate redback binary system within the previously-unidentified gamma-ray source 4FGL~J1910.7$-$5320 \citep{Au2022+J1910}. The discovery is a fruit of cross-matching the 4FGL-DR3 catalogue against sub-24 hour period optical variables in Catalina Real-Time Transient Surveys \citep{CSSPVC}.  4FGL~J1910.7$-$5320 was one of two spiders found in this way \citep[the other being PSR~J0955$-$3947;][]{Li2018+J0955}. SOAR/Goodman spectroscopy was also obtained, from which a sinusoidal radial velocity curve confirmed the binary nature of the system with an orbital period $P_B = 0.34847592$ days. The observed radial velocity amplitude, $K_{2,\text{obs}} = 218\pm8$km s$^{-1}$, is in line with what is seen in many redback systems, thus favoured as a redback candidate.
Independently of this optical discovery, we detected radio pulsations from this source as part of an ongoing survey for new pulsars in \textit{Fermi}-LAT sources \citep{TRAPUMSurv} being performed as part of the TRAnsients and Pulsars Using MeerKAT (TRAPUM) large survey project \citep{Stappers2016+TRAPUM}. This confirmed the redback prediction of \citet{Au2022+J1910}.

In this paper, we present the TRAPUM discovery of radio pulsations from the neutron star associated with 4FGL~J1910.7$-$5320 using the MeerKAT telescope. In \S \ref{observations} we describe the radio discovery and timing of the new pulsar, PSR~J1910$-$5320, as well as multi-band optical photometry obtained with ULTRACAM on the ESO New Technology Telescope. \S \ref{models} details the optical modelling of the optical light curves. In particular, we introduce a novel method to utilise values provided by radial velocity measurements made from optical spectroscopy. This modelling provides constraints on component masses, through the inclination and companion velocity, further confirming J1910 as a redback. \S\ref{DISCUSS} discusses the physical interpretation of our modelling, including an analysis of the impact of different gravity darkening prescriptions on the final results and an assessment of centre-of-light location where the absorption features are produced. A summary and conclusion is provided in \S\ref{Concs}.

\section{Observations}
\label{observations}
\subsection{Radio Discovery and Timing}
\label{radioobs}
In \citet{TRAPUMSurv}, we presented the first results from an ongoing survey being performed as part of the TRAPUM large survey project \citep{Stappers2016+TRAPUM} using the MeerKAT radio telescope \citep{Jonas2009+MeerKAT,Jonas2016+MeerKAT} to search for new pulsars in unassociated pulsar-like \textit{Fermi}-LAT sources. The survey presented therein consisted of two 10-minute observations of 79 sources from the 4FGL catalogue \citep{4FGL}, conducted using MeerKAT's $L$-band receiver (at observing frequencies between 856--1712\,MHz). This project has since been extended with a further two-pass survey (Thongmeearkom, T., et al., in prep.) being performed with the UHF receiver (544--1088\,MHz). Tied-array beams cover a larger solid angle at this lower frequency band, and so a small number of additional \textit{Fermi}-LAT sources whose localisation regions were too uncertain to cover in single observations at $L$-band were added to this UHF survey. One of these new sources was 4FGL~J1910.7$-$5320. 

TRAPUM observed this source on 2022 May 31, and detected highly significant radio pulsations with signal-to-noise ratio, S/N $\approx 380$. The signal had a spin period of 2.33\,ms and significant acceleration of $4.12 \pm 0.02$\,m\,s$^{-2}$ indicative of a millisecond pulsar in a short-period binary system. We used \texttt{SeeKAT}\footnote{\url{https://github.com/BezuidenhoutMC/SeeKAT}} \citep{SeeKAT} to localise this signal to a position less than 0.5$\arcsec$ from an optical star detected in the \textit{Gaia} DR3 \citep{Gaia,GaiaDR3} and Catalina Surveys Southern (CSS) periodic variable star catalogues \citep{CSSPVC}. The CSS catalogue lists this source as having a 16.8\,hr periodicity, with a double peaked light curve of 1.1\,mag amplitude. However, such a light curve is inconsistent with that of a pulsar binary companion, as the ellipsoidal modulation that gives rise to a double-peaked light curve has a maximum amplitude of around 0.3\,mag. However, folding the CSS data with half this period leaves a single-peaked light curve that is consistent with an irradiated binary pulsar companion star. Unknown to us at the time, this 8.4\,hr orbital period was independently confirmed by the optical spectroscopy presented in \citet{Au2022+J1910} through the measurement of Doppler-shifted spectral.

We therefore proceeded under the assumption that this star was indeed an irradiated redback counterpart to our newly-detected MSP, and used the CSS ephemeris to schedule follow-up timing observations with both MeerKAT and Murriyang, the Parkes 64m telescope, during the half of the orbit centered on the companion star's superior conjunction (i.e. orbital phases between 0.5 and 1.0) when the pulsar should not be eclipsed by wind from the companion.

Our timing campaign with MeerKAT consisted of 15 pseudo-logarithmically spaced observations between 2022 June 29 and 2022 September 29 with several observations on the first days (2022 June 29 and 2022 June 30) and increasing intervals between subsequent observations to facilitate phase connection. These observations each lasted 5\,min, and were taken using the Pulsar Timing User Supplied Instrument \citep[PTUSE,][]{Bailes2020+MeerTIME} with coherent de-dispersion. The first 8 observations were taken with MeerKAT's UHF receiver, the rest were performed at $L$-band. A second pseudo-logarithmic timing campaign began with Parkes on 2022 September 06 until 2023 March 25. These observations each lasted 1.5\,hr using the Ultra-wide-band Low (UWL) receiver \citep{Hobbs2020+UWL}, covering a frequency range from 0.7 to 4~GHz, with coherent de-dispersion. The resulting data were reduced using standard radio timing techniques, as described by \cite{TRAPUMSurv}, additional details will be provided elsewhere.

The resultant pulse times of arrival at the location of the radio telescope (ToAs) were analyzed using the \texttt{tempo} \citep{TEMPO} timing package . To model the motion of the radio telescope relative to the Solar System barycentre, we used the Jet Propulsion laboratory's DE421 Solar System ephemeris \citep{2009IPNPR.178C...1F}. To model the pulsar's
orbit, we used the BTX orbital model, which allows for the measurement of multiple orbital frequency derivatives. This is necessary because, like in most other redback systems, the ToAs revealed unpredictable deviations in the times of the pulsar's ascending node on the order of a few seconds, thought to be due to orbital period variations caused by variability of the companion star's gravitational quadrupole moment via the Applegate mechanism \citep{Applegate1992}. The parameters of the timing solution are presented in Table~\ref{t:timing}, where the numbers in parentheses indicated the 1-$\sigma$ uncertainties on the last digits of the nominal values. These parameters are presented in the Dynamic Barycentric time (TDB).

The determination of the timing solution was greatly assisted by previous knowledge of the orbital period (from CSS photometry) and the \textit{Gaia} astrometry, which was assumed for this solution.

\begin{table}
  \centering
  \caption{Timing solution for PSR~J1910$-$5320, obtained using the BTX orbital model. }
  \label{t:timing}
  \begin{tabular}{lc}
    \hline
    Parameter & Value \\
    \hline
    \multicolumn{2}{c}{\textit{Gaia} DR3 astrometry}\\
    \hline
    R.A., $\alpha$ (J2000) & $19^h10^m49\fs12053(1)$\\
    Decl., $\delta$ (J2000) & $-53\degr20\arcmin57\farcs1205(2)$\\
    Proper motion in $\alpha$, $\mu_{\alpha} \cos \delta$ (mas yr$^{-1}$) & $1.7\pm 0.2$\\
    Proper motion in $\delta$, $\mu_{\delta}$ (mas yr$^{-1}$) & $-6.8\pm 0.2$\\
    Parallax, $\varpi$ (mas) & $-0.42\pm0.26$\\
    Epoch of position measurement (MJD) & $57388.0$ \\
    \hline
    \multicolumn{2}{c}{Timing parameters}\\
    \hline
    Solar-system ephemeris & DE421\\
    Time scale & TDB \\
    Data span (MJD) & $59759.8$--$59978.5$\\
    Epoch of spin period measurement (MJD) & 59760\\
    Number of ToAs & 939 \\
    Residual rms ($\upmu$s) & 5.07 \\
    Reduced $\chi^2$ & 2.1 \\
    Spin frequency, $\nu$ (Hz) & $428.7490184657(3)$\\
    Spin-down rate, $\dot{\nu}$ (Hz s$^{-1}$) & $-6.80(7)\times10^{-15}$\\
    Dispersion measure, DM (pc cm$^{-3}$) & $24.42$\\
    Binary model & BTX\\
    Orbital frequency, $\nu_{\rm orb}$ (Hz) & $3.32132606(1)\times10^{-5}$\\
    First orbital frequency derivative, $\dot{\nu}_{\rm orb}$ (Hz s$^{-1}$) & $-3.45(4)\times10^{-18}$\\
    Second orbital frequency derivative, $\ddot{\nu}_{\rm orb}$ (Hz s$^{-2}$) & $1.85(4)\times10^{-25}$\\
    Projected semi-major axis, $x$ (lt s) & $0.969183(6)$\\
    Epoch of ascending node, $T_{\rm asc}$ (MJD) & $59759.9208124(3)$\\
    \hline
    \multicolumn{2}{c}{Derived parameters}\\
    \hline
    Spin period, $p$ (ms) & $2.33236685551(2)$\\
    Spin period derivative, $\dot{p}$ & $3.70(4)\times10^{-20}$\\
    Orbital period, $P_{\rm orb}$ (d) & $0.348477501(1)$\\
    Spin-down power, $\dot{E}$ (erg/s) & $1.15\times10^{35}$\\
    Surface magnetic field strength, $B_{\rm S}$ (G) & $3.0\times10^8$\\
    Light-cylinder magnetic field strength, $B_{\rm LC}$ (G) & $2.2\times10^5$\\
    \hline
  \end{tabular}
\end{table}

\subsection{Optical Photometry}
\label{opticalobs}
\begin{table*}
\caption{Time and phase coverage for ULTRACAM photometry obtained of J1910. The phase coverage, calculated with the timing ephemeris provided in Table \ref{t:timing}, corresponds with the phase axis of Figure \ref{fig:lcs}. The $g_s$ filter is split from the other two due to the exclusion of irreducible data for the 28/06/2022 night.}
\label{t:photometry}
\begin{tabular}{ccccl}
\hline 
Start Time (UTC)     & Observation length (hrs) & \multicolumn{2}{c}{Phase Coverage}                     \\
                                          &                          & $r_s$,$u_s$          & \multicolumn{1}{c}{$g_s$}       \\ \hline \hline
                     & \multicolumn{1}{l}{} & \multicolumn{1}{l}{}     & \multicolumn{1}{l}{} &                                 \\
29/06/2022 04:11:44          & 4.25              & 0.90 - 1.38          & \multicolumn{1}{c}{1.10 - 1.38} \\
\multicolumn{1}{l}{} & \multicolumn{1}{l}{} & \multicolumn{1}{l}{}     & \multicolumn{1}{l}{} &                                 \\
01/07/2022 01:31:58          & 5.0              & 0.30 - 0.89          & \multicolumn{1}{c}{0.30 - 0.89} \\
\multicolumn{1}{l}{} & \multicolumn{1}{l}{} & \multicolumn{1}{l}{}     & \multicolumn{1}{l}{} &                                 \\ \hline
\end{tabular}
\end{table*}
We obtained multi-band light curves of J1910 on two nights, 2022 June 28th and 30th, using the ULTRACAM high-speed multi-band photometer \citep{Dhillon2007}, mounted on the 3.50m New Technology Telescope (NTT) at the European Southern Observatory (ESO) La Silla, Chile. The times and length of these observations are provided in Table \ref{t:photometry}. ULTRACAM utilises 3 CCDs simultaneously, each using a different Super Sloan Digital Sky Survey (Super-SDSS) u$_s$g$_s$r$_s$i$_s$z$_s$ filter \citep{Dhillon2021}. For these observations CCDs 1, 2 and 3 used the r$_s$, g$_s$ and u$_s$ filters respectively. The data were taken under photometric conditions, with seeing varying between 1 - 1.5$''$. The observations were reduced using the HiPERCAM \citep{Dhillon2016, Dhillon2018} pipeline\footnote{\texttt{http://deneb.astro.warwick.ac.uk/phsaap/hipercam/docs/html/}}. Ensemble photometry \citep{Honeycutt1992} was used to calibrate the r$_s$ and g$_s$ bands. 12 nearby stars with known \textit{Gaia} magnitudes were chosen as reference apertures. In order to use the \textit{Gaia} magnitudes, they were transformed first into the SDSS prime r$'$ and g$'$ bands, then again into the corresponding HiPERCAM filters \citep[Appendix A]{Brown2022}. Due to a lack of \textit{Gaia} transform, and the unreliable transform between the HiPERCAM and SDSS filters, the u$'$ band was calibrated by using the instrumental zero point determined by observing the known SDSS standard PG1323-086D. After processing the data we were left with 3746 data points: 1608 and 1291 from the r$_s$ and g$_s$ bands respectively (20s exposures), and 530 from the u$_s$ (60s exposures). Co-addition of u$_s$ band exposures, maximising S/N, leaves fewer u$_s$datapoints relative to the other bands. The orbital phase of each point was calculated using the ephemeris given in Table \ref{t:timing}. Here the light curve phases have been folded as assumed in our ephemeris, with $\phi=0$ corresponding to the ascending node of the pulsar. Phases 0.25 and 0.75 therefore correspond with the companion's inferior and superior conjunctions respectively.
\begin{figure}
\centering

	\includegraphics[width = \linewidth]{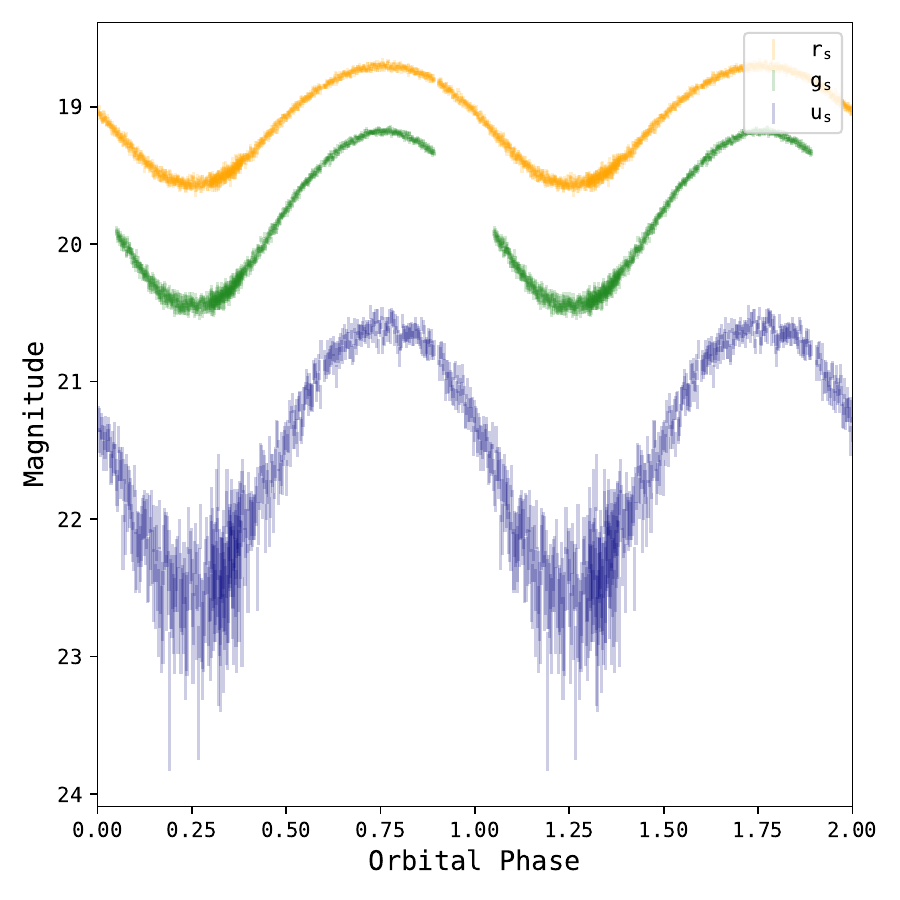}
	\caption{Phased light curves of J1910 in the r$_s$, g$_s$ and u$_s$ bands. The phasing is calculated according to the solution presented in Table \ref{t:timing}, with the companion's inferior and superior conjunctions occurring at phases 0.25 and 0.75 respectively. Two repeated cycles are shown for clarity. The gap in coverage in all bands around $\phi$ = 0.90 is due poor focus during the initial stages of the 2022/06/28 observing run. A larger portion of the g$_s$ band light curve is excluded due to irreducible artifacts in the data.  }
    \label{fig:lcs}
    
\end{figure}

\subsection{SOAR/Goodman Spectroscopy}
\label{spectroscopy}
The SOAR/Goodman spectroscopic data set for PSR~J1910$-$5320 is identical to that described in \citet{Au2022+J1910}. However, we found that the orbital ephemerides
inferred from these data show relatively modest but nevertheless quite statistically significant discrepancies with the ephemerides derived from pulsar timing. An investigation of these discrepancies led to the conclusion that a greater than expected degree of flexure was present in the previous SOAR/Goodman observations. Despite having calibration arc lamp observations continually interspersed throughout the object observations, and using night sky lines for an additional zeropoint correction, some residual effects of flexure remained. This could perhaps be associated with spatial flexure somewhere along the light path in the instrument, or instead with imperfect guiding that led to miscentering of the source in the slit.

\begin{table}
\centering
\caption{Updated radial velocities (RV) of PSR~J1910$-$5320 from SOAR for both the full spectrum and targeting just the Mg$\beta$ triplet.}
\label{t:rv}
\begin{tabular}{lllll}
    \hline 
     & \multicolumn{2}{c}{Full spectrum} & \multicolumn{2}{c}{Mg$\beta$} \\ 
    
    BJD    & \multicolumn{1}{c}{RV} & \multicolumn{1}{c}{$\Delta$RV} & \multicolumn{1}{c}{RV} & \multicolumn{1}{c}{$\Delta$RV} \\ 
    (days) & \multicolumn{2}{c}{(km s$^{-1}$)} & \multicolumn{2}{c}{(km s$^{-1}$)}  \\\hline \hline
     &               &    & &            \\
    59679.32228	& -17.8	& 21.7 &	-28.9 &	25.3             \\
    59679.34012	& -58.4	& 20.6 & -72.4 & 26.5 \\
    59679.35806	& -141.8 & 18 & -164.2 & 26.3 \\
    59680.32745	& 80.7 & 17.5 & 82.1 & 21.1 \\
    59680.34496	& 67.5 & 19.1 & 79 & 21.4 \\
    59680.36458	& 16.1 & 16.3 & -7.5 & 19.3 \\
    59700.30193	& -218.9 & 15.7 & -219.2 & 19.7 \\
    59700.31948	& -243.2 & 16.3	& -240.5 & 20.4 \\
    59700.33924	& -206.4 & 18.2	& -240.7 & 22.1 \\
    59722.16463	& 76.9 & 14.8 & 98.8 & 24 \\
    59722.18212	& 42.6 & 15.5 & 43.8 & 18.5 \\
    59724.34289	& -187.2 & 19.1	& -239.4 & 32.6 \\
    59724.36042	& -188.9 & 20.7	& -203.4 & 32.4 \\
    59724.38436	& -226.2 & 33.3	& -253.4 & 40.6 \\
    59740.19389	& 150.5 & 22.2 & 160.2 & 33.4 \\
    59740.21175	& 146.4	& 28 & 164.2 & 33 \\
    59740.30736	& -20 & 16.1 & 3.4 & 22.1 \\
    59740.32483	& -60.9	& 17.1 & -86.1 & 23 \\
    59740.38604	& -149.9 & 19.2	& -161.6 & 21.7 \\
    59740.40352	& -228.8 & 23.8	& -261.7 & 27.9 \\
     &               &    &       &             \\ \hline
\end{tabular}
\end{table}
Therefore, we have re-derived the PSR~J1910$-$5320 radial velocities through a process that differs in some details from the method used in \citet{Au2022+J1910}. To improve the wavelength zeropoint corrections, we use the {\tt TelFit} code \citep{Gullikson2014} to generate a telluric absorption spectrum based on the airmass, the local humidity, pressure, and temperature, and the 3-hour Global Data Assimilation System atmospheric model closest in time to each object spectrum. This model spectrum, smoothed to the resolution of the SOAR data and binned to the same pixel scale, is then fit to the object spectrum in the region of the Fraunhofer A band (7580-7700 \AA) to determine the wavelength zeropoint correction. While other telluric features are also present in some spectra, this is the only telluric feature measurable in essentially all usable spectra, even those of low signal-to-noise, so we restrict the fit to this feature. Comparisons over a number of datasets show that the corrections from this method are generally similar to, but sometimes more accurate than, those from the night sky lines.

We also re-fit the object radial velocities with {\tt RVSpecFit} \citep{Koposov2011+RVspecfit,Koposov2019+RVspecfit}, using a library of PHOENIX synthetic templates \citep{PHOENIX} of varying metallicity, temperature, surface gravity, [$\alpha$/Fe] abundance, as well as allowing for rotation.
As described in \S\ref{Intro} companion surface heating complicates the measurement; the inferred velocity does not necessarily track the true centre-of-mass velocity, rather the centre of light associated with a specific line. This is clearly reflected by the differing K$_2$ amplitudes determined in \citet{Au2022+J1910}, and updated here in Table \ref{t:rv}, when considering the full spectrum versus only the Mg$\beta$ triplet \citep[a similar treatment is given in][]{Linares2018}.

Hence for each spectrum we performed two fits: the first over the entire range of the optical spectrum with measurable absorption lines (4000–6800 \AA) and the second solely in the region of the Mg$\beta$ line. Overall, the inferred velocities from this method are consistent with those obtained from cross-correlation with an appropriate template over a comparable wavelength range. 

\section{Photometric modelling}
\label{models}
The optical light curve modelling performed here utilised the binary stellar synthesis code \texttt{Icarus} \citep{Breton2012}, with some novel modifications. As such, the procedure followed is comparable, though not identical, to the modelling performed in similar analyses \citep{Breton2013, Draghis2019, Stringer2021, Kennedy2022, DMS2023}. Here the specific procedure and priors used for this system will be described  \citep[see][for a more in-depth description of \texttt{Icarus}]{Breton2012}.

\subsection{Surface heating models}
\label{heating}
Compared to previous uses of \texttt{Icarus}, not limited to those cited above, here we have amended the gravity darkening prescription applied to the companion's surface. Previously the temperature of companion surface element $i$, $T_i$, \textbf{before} irradiation was calculated as
\begin{equation}
    T_{i} = T_\text{base} \left(\frac{g_i}{g_\text{pole}}\right)^\beta 
\end{equation}
where $T_\text{base}$ is the \texttt{Icarus} input parameter specifying the temperature at the pole of the star, $g_i$ is the surface gravity at surface element $i$, $g_\text{pole}$ is the surface gravity at the pole of the star and $\beta$ is the gravity darkening coefficient. This equation still applies here, though its deployment differs in two significant ways:
\begin{enumerate}
    \item We assume the companion's atmosphere heat transfer \textit{close to the surface} is radiative, as opposed to convective. A radiative gravity darkening coefficient ($\beta$) of 0.25 was used, as opposed to the usual 0.08 used for a convective atmosphere \citep{Breton2013}. 
    \item We include the option to apply gravity darkening after irradiation and heat redistribution on the heated companion surface. This differs from the previously standard \texttt{Icarus} behaviour to gravity darken the base (singular temperature) companion surface before heating effects are considered.
\end{enumerate}

We found that these changes improve our model fits substantially and are physically motivated by a number of new insights we gained on the stellar physics. For the first assumption, following \citet{Zilles2020}, we expect the inner photosphere of the companion to be convective where the Schwarzschild criterion is satisfied, and radiative toward the surface. Therefore the gravity darkening prescription for the photosphere surface should follow the radiative law. \citet{L&R2012} also demonstrated that tidally distorted low-mass, convective stars should in fact present gravity darkening coefficients in the interval $[0.20,0.25]$, with spider-like companions being at the upper end of this range.

Though this latter work does not include the effects of irradiation, there is a strong possibility that the irradiation impinging onto J1910's companion, and other spider companions, leads to deep heating of their photosphere. This is in contrast to our previous application of gravity darkening before irradiation, which implicitly assumed it was only superficial. The fact that spectral lines in these systems are generally absorption features (except for a few emission line features which are likely connected to outflowing material) indicates that irradiation is deposited deep enough for no substantial thermal inversion to occur as is seen in the case of cataclysmic variables where the shallow heating is caused by UV photons from a hot white dwarf. It then follows that the irradiating flux should be considered a fundamental aspect of the surface temperature profile, and as such gravity darkened along with the rest. As the exact depth of the heating in J1910 is unclear and a full theoretical treatment of its effect on gravity darkening not available at the moment, we opted to test both pre- and post-irradiation gravity darkening models for completeness.

The parameters fit for using \texttt{Icarus} depended on the surface heating model applied. The most basic model, direct heating (DH), applies symmetrical irradiation onto the companion's inner face, locked toward the pulsar. The parameters fit for this model constitute our fundamental set: the systemic velocity $\gamma$, the interstellar reddening E(B-V), the system inclination $i$, the Roche-lobe filling factor $f_\text{RL}^*$\footnote{$^*$ Calculated as $\frac{r_N}{r_L1}$, where $r_N$ is the distance from the companion's barycentre to its nose, and $r_L1$ is the distance from the barycentre to the L1 point.}, the base and irradiating temperatures $T_\text{base}$ and $T_\text{irr}$, the distance $d$ and the projected radial velocity amplitude of the companion K$_2$. 

Heat redistribution across the stellar surface was also considered, as set out in \citet{Voisin2020}. For an irradiated companion face with temperature differences between the dayside and nightside, diffusion of heat from the irradiated face can be expected. In our models this is accounted for by adding two parameters to our `fundamental' parameter set: $\kappa$, which parameterises the amplitude of the diffusion effect, and $\Gamma$, which governs the temperature dependence of the diffusion \citep{Stringer2021}. In this case, we have elected not to include $\Gamma$. Trial fits including it regularly found very little constraint on it, and those without obtained a better Bayesian evidence without significant effect on other parameters. 

Heat redistribution models can also account for asymmetrical light curves, found for a number of spiders \citep{Stappers2001, R&S2016, Linares2018, K&R2020, Romani2021, Stringer2021}, whereby light curves at not symmetric between the half orbits centered on the companion's ascending and descending nodes. Three main approaches have usually been implemented to account for this:
\begin{enumerate}
    \item A convective wind following a certain latitudinal profile, with strength parameterized by $C_\text{amp}$.
    \item A surface hot/cold spot with fitted temperature, size and position \citep[e.g.][]{Clark2021}.
    \item Re-distribution of irradiating flux by an extended, swept-back intra-binary shock \citep{R&S2016} and/or magnetic ducting \citep{R&S2017}.
\end{enumerate}

These models account for asymmetry by shifting or adding flux onto one side of the companion's inner face, such that more/less flux is seen at ingress/egress to the companion's superior conjunction. In this work we have focused on using diffusion and convection (D+C) models to redistribute heat across the companion's surface. Whilst hot spots are well-supported in literature and physically \citep{R&S2017}, in the present case spot models invariably placed the spot, given the modelled inclination, largely out of sight on the companion's surface at all orbital phases. We took this as an indication that a spot model was not suitable for J1910. 

The parameters set for each model were sampled and constrained by channeling \texttt{Icarus} through \texttt{dynesty} \citep{dynesty}, a Python implementation of a dynamic nested sampling Bayesian parameter and evidence estimation algorithm \citep{Skilling2004, MULTINEST1, MULTINEST2, MULTINEST3}. Nested sampling algorithms provide the Bayesian evidence of a model, $Z$, a useful advantage over a classic Monte Carlo Markov Chain (MCMC) algorithms. Allowing for the calculation of the Bayes factor,
\begin{equation*}
    B_{1,2} = \frac{Z_1}{Z_2} \,,
\end{equation*}
between two models enables one to determine which is favoured; $B_{1,2}$ > 1 suggests model 1 is preferable, whereas $B_{1,2}$ < 1 would prefer model 2 \citep{Jeffreys1939}. The basic procedure on a given iteration of the nested sampler, using only the optical photometry, first selects a set of samples from the provided priors, passing them into \texttt{Icarus}. The likelihood is calculated from the $\chi^2$ fit of the observed photometry and the simulated light curves generated given sampled parameters.

\subsection{Priors}
\label{Priors}
Careful consideration must be given to the choice of priors for our models and, where possible, they should be strongly motivated by physical or geometric constraints or, in the case of K$_2$, the use of complementary independent data (\citet{Au2022+J1910}, \S\ref{RV fit}). The main priors used here were as follows:
\begin{itemize}
    \item A Gaussian prior applied to E(B-V), centered on the reddening provided by the dust maps of \citet{S&F2011}: 0.0596 $\pm$ 0.0033.
    \item A simple $\sin(i)$ prior applied to $i$, corresponding to an isotropic distribution of orbital angular momentum vectors.
    \item A distance prior constructed using the same procedure as in \citet{Clark2021} and \citet{Kennedy2022}. This combines the expected density of Galactic MSPs along the line of sight to J1910 \citep{Levin2013}, the transverse velocity distribution for binary MSPs in the ATNF Pulsar catalogue \citep{ATNFcat} and the \textit{Gaia} DR3 parallax \citep{GaiaDR3}. Additional constraint can be provided by the DM inferred from radio timing using the Galactic electron density model \citet[YMW16,]{Yao2017}. In the present case, we have opted not to employ it. The DM distance is not equally reliable for all lines of sight, and the distance inferred from the DM ($0.92 \pm 0.49$ kpc) is much smaller, and less reliable, than that from the \textit{Gaia} parallax ($6.8 \pm 3.9$ kpc). \citet{Yao2017} themselves compiled a list of pulsars with independent distance measurements both underestimated and overestimated by their model, therefore an underestimation from it for J1910 is not entirely unexpected. 
\end{itemize}

\subsection{Spectroscopic K$_2$ constraint} 
\label{RV fit}
Given the very high-precision timing measurement of the pulsar's projected velocity amplitude, any measurement of the companion's K$_2$ determines the mass ratio $q$, and then provides a constraint on the masses via the mass function of the system. K$_2$ is typically measured from the Doppler motion of absorption lines over the orbit, to which a centre-of-light correction must be applied.

Previous iterations of \texttt{Icarus} have allowed for the incorporation of spectroscopic data in various ways. \citet{Clark2021} calculated an average of companion surface element velocities (simulated as part of \texttt{Icarus}) over the orbit, weighted by their flux to compensate for centre-of-light effects in an approximate manner. The resulting model radial velocities were subtracted from the observed radial velocities, and the overall model penalised according to the resulting likelihood. \citet{Kennedy2022} used a self-consistent procedure, where observed spectra were directly fitted to simulated spectra generated by \texttt{Icarus} from \texttt{ATLAS9} \citep{ATLAS9} atmosphere grids to produce a likelihood. This method intrinsically overcomes the centre-of-light issue, as irradiation is implicit in the generated model spectra. There is, however, a significant computational cost associated with simulating full model spectra and a potential risk for the fitting to try and reproduce features of the spectrum which are not well accounted for by the atmosphere model.

In this work a middle ground between the two methods described above was used, balancing adequate simulation of the spectra with computational expense. As with the self-consistent spectroscopy modelling of \citet{Kennedy2022}, here \texttt{Icarus} is used to simulate spectra for each sample. However, these spectra were not directly compared with their observed counterparts, rather the radial velocities of the models were determined and compared to their experimental analogues. Specifically narrow, and thus inexpensive, spectra centred around the 5183 \AA\, Mg$\beta$ triplet were generated for each orbital phase covered by the SOAR/Goodman dataset. The radial velocity for each phase was determined by cross correlating the spectrum at a reference orbital phase (chosen to be that showing the strongest line feature), thus providing a relative projected radial velocity curve. The likelihood between the observed and modelled radial velocities was then incorporated into the fitting procedure.

\section{Modelling results}
\label{DISCUSS}
\begin{table*}
\begin{tabular}{lllllll}
\hline
                     & \multicolumn{2}{c}{DH}                             & \multicolumn{2}{c}{D+C}                             \\
Param                & \multicolumn{1}{c}{Pre} & \multicolumn{1}{c}{Post} & \multicolumn{1}{c}{Pre} & \multicolumn{1}{c}{Post}  \\ \hline \hline
\multicolumn{1}{c}{} & \multicolumn{6}{c}{Icarus Parameters}                                                                                                                        \\\hline
                     &                         &                          &                         &                           \\
Interstellar reddening, E(B-V)               &    $ 0.060^{+0.003}_{-0.003} $                     &    $ 0.060^{+0.003}_{-0.003} $                      &  $ 0.060^{+0.003}_{-0.003} $                       & $ 0.060^{+0.003}_{-0.003} $                         \\
                     &                         &                          &                         &                                                  \\
Inclination angle, $i$ (deg)                   &    $54^{+3}_{-3}$                     &   $52^{+2}_{-2}$                        & $46^{+1}_{-1}$                        & $45^{+1}_{-1}$                                      \\
                     &                         &                          &                         &                                            \\
Roche-lobe filling factor, $f_{\rm RL}$             &    $ 0.818^{+0.007}_{-0.008} $                     &   $ 0.762^{+0.005}_{-0.005} $                        & $ 0.838^{+0.008}_{-0.010} $                        & $ 0.782^{+0.005}_{-0.005} $                           \\
                     &                         &                          &                         &                         \\
Base temperature $T_{\rm base}$ (K)               &     $5460^{+40}_{-40}$                    & $5360^{+42}_{-50}$                         & $5310^{+40}_{-40}$                        &  $5200^{+40}_{-40}$                  \\
                     &                         &                          &                         &                        \\
Irradiating temperature $T_{\rm irr}$ (K)               &     $6060^{+60}_{-60}$                    & $6700^{+70}_{-70}$                         & $6240^{+80}_{-70}$                        &  $6760^{+70}_{-70}$           \\
                     &                         &                          &                         &                      \\
Distance, $d$ (kpc)                   &     $4.2^{+0.2}_{-0.2}$                    & $4.0^{+0.1}_{-0.1}$                         & $4.8^{+0.2}_{-0.2}$                        &  $4.43^{+0.2}_{-0.1}$          \\
                     &                         &                          &                         &                       \\
Companion radial velocity amplitude, $K_2$ (km s$^{-1}$)               &     $206^{+7}_{-7}$                    &  $197^{+8}_{-7}$                        & $216^{+8}_{-7}$                        &  $200^{+8}_{-8}$        \\
                     &                         &                          &                         &                     \\ \hline
\multicolumn{1}{c}{} & \multicolumn{6}{c}{Heating parameters}                                                                                                                       \\
\hline
                     &                         &                          &                         &                      \\
Diffusion coefficient, $\kappa$             &  -                       &  -                        &  $900^{+1200}_{-600}$                       & $80^{+100}_{-50}$                        \\
                     &                         &                          &                         &                       \\
Convection amplitude $C_\text{amp}$       &  -                       &  -                        & $-1700^{+100}_{-100}$                         &  $-1760^{+60}_{-60}$         \\
                     &                         &                          &                         &                        \\ \hline
\multicolumn{1}{c}{} & \multicolumn{6}{c}{Derived parameters}                                                                                                                       \\
 \hline
                     &                         &                          &                         &                        \\
Mass ratio, $q$                    &   $3.4^{+0.1}_{-0.1}$                      & $3.2^{+0.1}_{-0.1}$                         & $3.6^{+0.1}_{-0.1}$                        & $3.3^{+0.2}_{-0.2}$                         &                         &                          \\
                     &                         &                          &                         &                       \\
Pulsar mass, $M_P$ (M$_\odot$)                &    $ 1.0^{+0.1}_{-0.1} $                     &  $ 1.0^{+0.1}_{-0.1} $                        &  $ 1.6^{+0.2}_{-0.2} $                       & $ 1.4^{+0.2}_{-0.2} $                 \\
                     &                         &                          &                         &                         \\
Companion mass, $M_C$ (M$_\odot$)               &     $ 0.29^{+0.03}_{-0.04} $                    &  $ 0.29^{+0.03}_{-0.03} $                        & $ 0.45^{+0.04}_{-0.04} $                        & $ 0.43^{+0.04}_{-0.03} $                 \\
                     &                         &                          &                         &                \\
Observed temp. at companion superior conjunction, $T_\text{S}$ (K)      &  $6170^{+50}_{-50}$ & $6110^{+50}_{-60}$                      &   $6060^{+40}_{-40}$                                                &  $5940^{+50}_{-60}$           \\
                     &                         &                          &                         &                      \\
Observed temp. at companion inferior conjunction, $T_\text{I}$ (K)    &  $5140^{+30}_{-30}$                      &   $5100^{+30}_{-40}$                      &  $5070^{30}_{-30}$                       &  $4990^{+40}_{-40}$                \\
                     &                         &                          &                         &                      \\
Day-side temperature, $T_\text{day}$ (K)      &  $6350^{+40}_{-40}$ & $6300^{+50}_{-50}$                      &   $6340^{+40}_{-40}$                                                &  $6220^{+50}_{-60}$           \\
                     &                         &                          &                         &                      \\
Night-side temperature, $T_\text{night}$ (K)    &  $5050^{+40}_{-40}$                      &   $5000^{+40}_{-40}$                      &  $4930^{+30}_{-30}$                       &  $4840^{+40}_{-40}$                \\
                     &                         &                          &                         &                      \\
Volume-averaged filling factor, $f_\text{V}$    &  $0.947^{+0.004}_{-0.004}$                      &   $0.911^{+0.004}_{-0.004}$                      &  $0.957^{+0.004}_{-0.006}$                       &  $0.924^{+0.004}_{-0.003}$                \\
                     &                         &                          &                         &                       \\ 
Irradiation efficiency, $\epsilon$    &  $ 0.22^{+0.03}_{-0.03} $                       & $ 0.32^{+0.03}_{-0.03} $                         & $ 0.34^{+0.03}_{-0.03} $                        & $ 0.42^{+0.04}_{-0.04} $       \\
                     &                         &                          &                         &                       \\ 
Transverse velocity, $v_T$ (km s$^{-1}$)    &  $ 140^{+6}_{-7} $                       & $ 132^{+5}_{-5} $                         & $ 158^{+6}_{-5} $                        & $ 144^{+5}_{-5} $       \\
                     &                         &                          &                         &                      \\ \hline                    
\multicolumn{1}{c}{} & \multicolumn{6}{c}{Fit Statistics}                                                                                                                           \\
\hline
                     &                         &                          &                         &                      \\
Photometry $\chi^2$ (3446 datapoints)       &  4889.585                   & 5106.263                       & 3611.385                     &  3637.874                       \\
                     &                         &                          &                         &                      \\
Radial velocity $\chi^2$ (20 datapoints)      &  29.144                   & 29.830                       & 30.456                     &  29.132                       \\
                     &                         &                          &                         &                      \\
log-Evidence, $Z$                    & $-2366.3\pm0.2$                        & $-2442.1\pm0.2$                         & $-1851.6\pm0.2$                        & $-1865.8\pm0.2$   \\
                     &                         &                          &                         &                          &                         &                          \\
log Bayes Factor vs. Convective                   & $119$                        & $43$                         & $634$                        & $619$  \\
                     &                         &                          &                         &  \\
log Bayes Factor vs. Radiative                   & $0$                        & $-76$                         & $514$                        & $501$  \\
                     &                         &                          &                         &  \\\hline                    
\end{tabular}
\caption{Posterior parameter results from photometry and radial velocity curve fitting. Results are split into the two key models used: direct heating (DH), which employs no heat redistribution, and diffusion + convection (D+C). These are subsequently split by the gravity darkening prescription, pre- or post-irradiation (including heat redistribution effects). Note the \texttt{Icarus} parameters $T_B$ and $T_I$ do not reflect the `true' physical conditions on the companion's surface. Rather, $T_\text{S}$ and $T_\text{I}$ average the visible surface element temperatures at the companion's superior and inferior conjunctions respectively. $T_\text{day}$ and $T_\text{night}$ then provide the intrinsic temperatures of the day and night sides, again averaging surface element temperatures assuming an edge on inclination. Quoted uncertainties correspond to 68\% confidence intervals. The (log) Bayesian evidence ($\ln Z$) produced by \texttt{dynesty} is used to calculate the Bayes factors between a given model and a reference one, chosen to be DH with pre-irradiation gravity darkening.}
\label{t:results}
\end{table*}

\begin{figure*}
\centering
	\includegraphics[width = \linewidth]{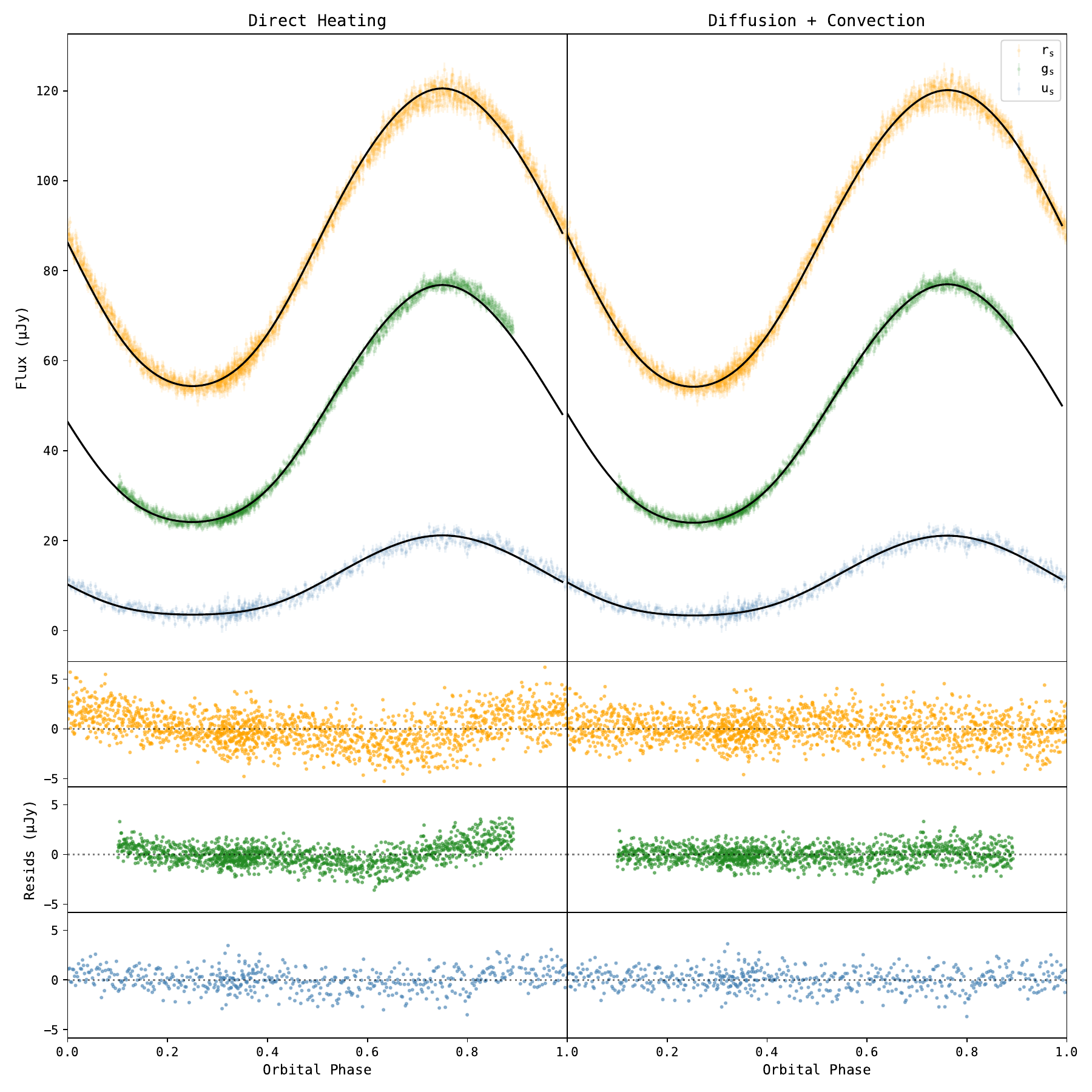}
	\caption{Photometry fits produced by post-irradiation gravity darkening models presented in Table \ref{t:results}. The maximum a posteriori likelihood models have been selected. Left shows the direct heating model (DH), while right is diffusion + convection (D+C). The light curve data (Fig \ref{fig:lcs}) are shown in the corresponding colours, with model fits overlaid in black. Residuals for each band are shown below. Clearly visible between the two panels is the improvement in the residuals with the introduction of diffusion + convection to address the asymmetry in the light curve.}
	\label{f:models}
\end{figure*}

\begin{figure*}
\centering
    \includegraphics[width = \linewidth]{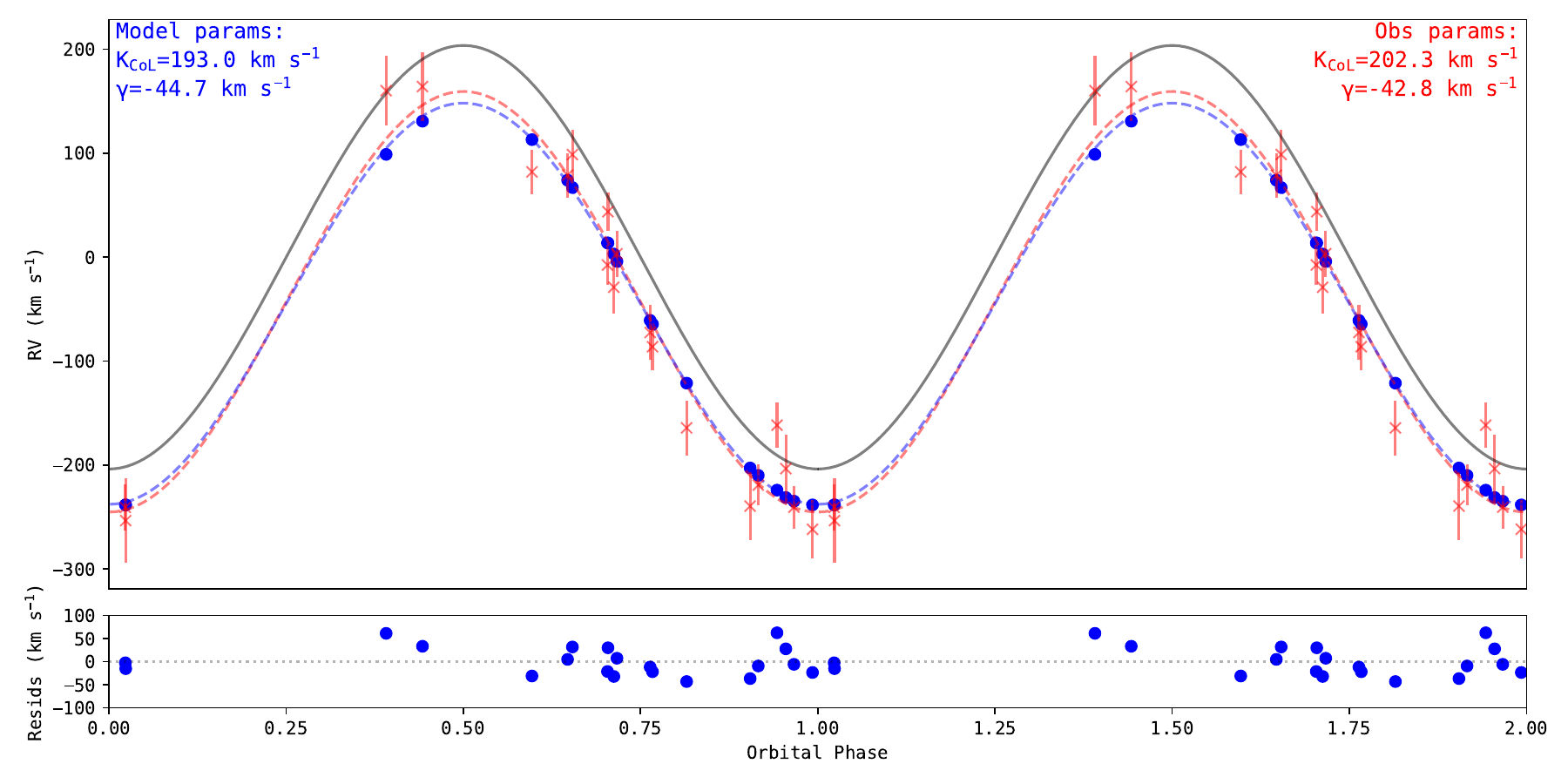}
	\caption{Mg$\beta$ radial velocity curve fit for post-irradiation gravity darkening diffusion + convection model. The top panel shows our model radial velocity points, blue, against the observed curve, red. The corresponding dashed lines are sinusoidal fits through each set, giving the parameters in the top corners. The grey solid line is the centre-of-mass radial velocity curve, using the underlying $K_2$ for the best-fit model. Point wise residuals between the model and observed points are shown in the bottom panel.}
\label{fig:specRV}
\end{figure*}

Table \ref{t:results} contains the results for the models considered and discussed above. These are split by heating model (DH or D+C) and subsequently by the prescription used to apply gravity darkening (pre- vs post-irradiation and heat redistribution effects). In both heating models a consistent trend emerges: post-irradiation gravity darkening finds a smaller projected companion velocity $K_2$. Before dissecting the differences between the pre- and post-irradiation gravity darkening, we can first get an overall picture of the parameters determined for this newly modelled system.

The DH models are presented for completeness, they do not constitute favourable models. The left hand panels of \ref{f:models} show the post-irradiation gravity darkened DH model fit to the data. Paying attention to the residuals, the asymmetry in the light curve becomes clear. The model both overestimates the flux at the ingress to the optical maximum and underestimates the flux at the egress. The 12 reference stars used in ensemble photometry show no consistent excess corresponding to these orbital phases, thus it is safe to assume this is intrinsic light curve asymmetry. As such, the extremely low pulsar masses determined for both DH models can be safely discarded.

Our D+C models are much better than DH models at capturing the behaviour of the data and can account well for the asymmetry. The inferred $C_\text{amp}$ implies a convective surface wind blowing in the direction of the companion's rotation, and thus depositing heat towards the companion's leading edge. The improvement in the fit is reflected in the statistics provided in Table \ref{t:results}. The underlying reasons for changes in parameter values are far from trivial to pin down, but notable is a shift in $i$ between the DH and D+C models, which implies a different inferred pulsar mass. Given a DH model will struggle to fit the amplitude of a asymmetric light curve it is unsurprising that $i$, which directly modulates the amplitude of an optical light curve, will be affected once heat redistribution is incorporated.

When compared with similar \texttt{Icarus} modelling results involving asymmetric heat redistribution, J1910 is the only redback in which the heat is transferred to the leading edge (i.e. excess flux near descending node of the companion). PSRs J2215+5135 \citep{Voisin2020}, J1227-4853 and J1023+0038 \citep{Stringer2021} all show excess flux toward the trailing edge of the light curve (i.e. excess flux near ascending node of the companion). Though we draw no major assertions from it, J1910 marks an notable departure from previously modelled redbacks.

\subsection{Overall constraints}
\label{Constraints}
Considering now only the D+C models, a number of parameters agree across both gravity darkening options. The inclination remains consistent around 45$^\circ$, with both models agreeing within their respective $68\%$ confidence interval. The irradiating temperatures in both models are consistently above 6000 K. More importantly, both models find average temperatures -- where the temperatures across the visible surface are averaged in their 4th power, i.e. according to their bolometric luminosity, and weighted by the projected surface area -- at the observed superior and inferior conjunctions that agree within their $68\%$ confidence intervals. This means that both models essentially reproduce the same colours in these parts of the light curves. From the lowest and highest points of the 68$\%$ confidence regions, we find $4950 < T_I < 5100$. This is slightly lower than our expectation from the broadband spectral energy distribution (SED) but within the allowed uncertainty \citep{Au2022+J1910}. $C_\text{amp}$ also agrees well for both which is expected given this parameters controls the asymmetry in the light curve.

\begin{figure}
\centering
    \includegraphics[width = \linewidth]{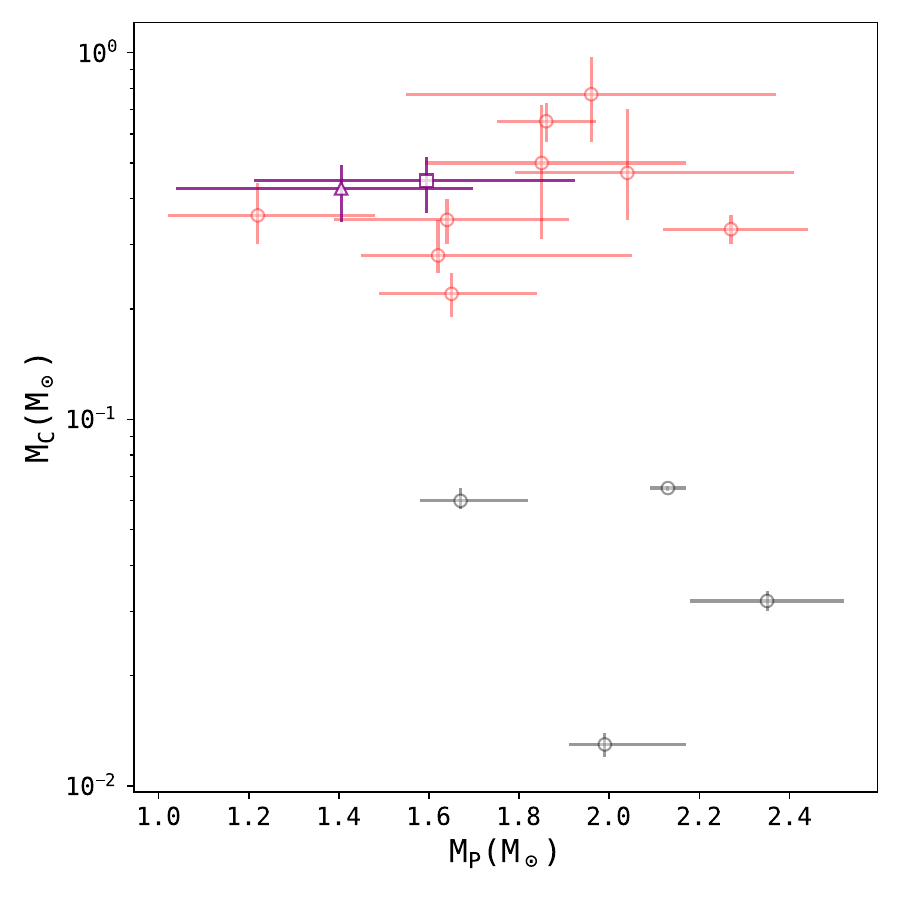}
	\caption{Companion ($M_C$) and pulsar ($M_P$) masses for a selection of redbacks (red) and black widows (black). Pre and post gravity darkening diffusion + convection models for J1910 are shown by the purple square and triangle respectively. Spider mass demographics sourced from \citet{Strader2019, Romani2021, Romani2022, Nieder2020, Clark2021, Kennedy2022} and references therein.}
    \label{fig:masses}
\end{figure}

Several parameters are not consistent between models, though we can still produce `ballpark' educated guesses at their values. The filling factors do vary between the models, but not over a large range, with both implying a significantly under-filling companion. Moving from the \texttt{Icarus} parameter $f_\text{RL}$ to the volume averaged filling factor we find an even smaller interval. Though significantly higher than the \texttt{Icarus} parameter $f_\text{RL}$, these should still be interpreted as under-filling, particularly the post-irradiation gravity darkening case. 

A key aim of light curve modelling in spider systems is to constrain the pulsar mass. Figure \ref{fig:masses} shows a collection of spider mass measurements, with the masses determined for our D+C models shown in purple; the square and triangle denoting the pre- and post-irradiation gravity darkening models, respectively. In this case we get a two moderate masses depending on the model chosen - none threaten the upper end observed pulsar masses and thus are useful to constrain the dense matter equation of state on their own.

\citet{Linares2019} collated a number of `super-massive' neutron star mass measurements. The quality of our measurement is at a similar level to other spiders in this sample - especially those without independent constraints on either the inclination or companion mass. For example, PSR~B1957+20's recently updated mass constraint uses $\gamma$-ray eclipsing to provide hard constraints on the inclination \citep{Clark2023b}. We do not reach the same mass precision as \citet{Kennedy2022} or \citet{Romani2021}, where the full, high S/N spectroscopy has been used in constraining the model. The high precision masses determined for relativistic NS-NS binaries, utilising post-Keplerian parameters measured through pulsar timing, out perform the measurement here as do measurements for NS-WD binaries (see \citet{Lattimer2012}). The systematics inherent to spider light curve modelling, namely the reliance on inferring a heating model for the surface, somewhat limit the precision we can expect to achieve. As these systematics are chiefly driven by irradiation, they are typically assumed to be lessened in redbacks when compared with black widows \citep{Strader2019}. However as J1910 is an irradiation-driven redback, significant surface heating must be accounted for. The precision of J1910's mass measurement, as well as other irradiation-dominated spiders, is closely tied to our understanding of the irradiation in these systems (see \citet{R&S2016,R&S2017,Voisin2020,Zilles2020}). In addition to full spectroscopy modelling, using high signal-to-noise spectra, and independent constraints would allow for a more precise mass measurement. Unfortunately here the inferred inclination is too low for a $\gamma$-ray eclipse, removing one independent constraint we might appeal to \citep{Clark2023b}. 

\begin{figure*}
\centering
    \includegraphics[width = \linewidth]{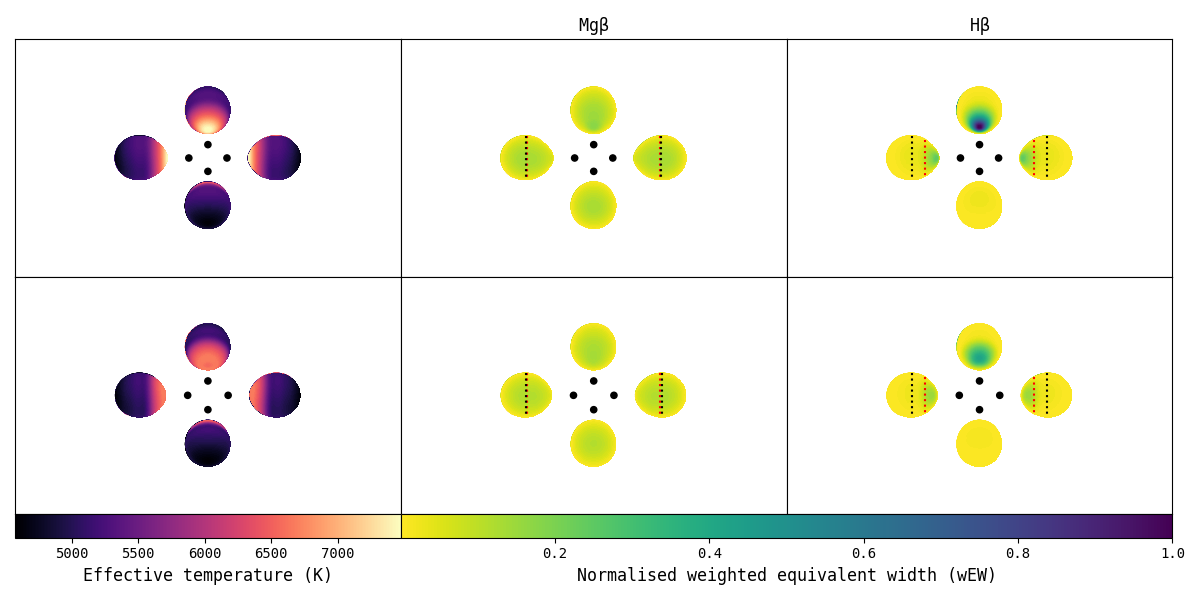}
	\caption{Surface maps for pre (top) and post (bottom) irradiation gravity darkening diffusion  + convection models. The leftmost plot shows the surface temperature over the companion surface. The two plots on the right show the normalised flux weighted equivalent width (wEW) from each surface element (see (\ref{eq:wEW})). These are split into the Mg$\beta$ triplet, which corresponds with our radial velocity curve, and the H$\beta$ feature. In the picture of \citet{Linares2018} these track the companion nightside and dayside respectively. The dashed lines on the wEW maps indicate the centre of mass (black) and centre of light (red) positions for the given line. Recall that a centre of light towards the companion's nose should correspond with a lower radial velocity determined for that line than the true centre-of-mass radial velocity (sampled by \texttt{Icarus} and used to calculate $M_P$).}
\label{fig:surface}
\end{figure*}

\subsection{Gravity darkening}
\label{GD}
Changing the gravity darkening prescription, as detailed in \S\ref{heating}, has a notable effect on the inferred pulsar mass in J1910; a higher $M_P$ for pre-irradiation gravity darkening, and a lower one for post-irradiation. Masses in the system are not directly fitted for; they are derived from other parameters, and most specifically from $i$ and $K_2$. Given the high-precision binary mass function determined from the radio timing, the pulsar mass should roughly scale with the cube of the companion's centre-of-mass velocity and inversely with the cube of $\sin i$. As $i$ does not change significantly between the two prescriptions, $K_2$ must primarily drive the variation in pulsar mass. From the ratio $K_2$ between the two models, we would expect a $\sim 25\%$ change in mass, while the actual difference is $\sim 15\%$. This implies that the changes cannot be entirely treated in isolation and that correlations between these two key parameters, and other ones from the model, contribute to dictating the masses.

Separately, we also observe that going from the pre-irradiation to the post-irradiation prescription causes the inferred values of $f_\text{RL}$, $T_\text{irr}$ and $q$ to decrease, and $T_\text{irr}$ to increase. Allowing for the irradiated face of the companion to be gravity darkened changes the balance between the irradiating flux and the star's size (mediated by $f_\text{RL}$). The exact interplay between these parameters is difficult to disentangle and, while we cannot summarise it with a single effect, we can suggest a few correlations.

Changing the gravity darkening prescription naturally changes the heating pattern on the companion's surface. Temperature maps produced post-irradiation gravity darkening appears to shift heat, and thus flux, away from the centre of the irradiated face and toward the sides of the companion. This will shift the centre of light for any spectral lines, in our case the Mg$\beta$ triplet, toward the centre-of-mass. Therefore, to match the observed line velocities, the sampled centre-of-mass $K_2$ must decrease to compensate. This effect is explored further in \S\ref{s:CoL}. This shifting of flux to the sides is likely linked to the smaller diffusion coefficient $\kappa$ found for the post-irradiation gravity darkening model.

$K_2$ directly constrains the mass ratio, which in turn changes the size of the companion's Roche lobe. Decreasing the companion's size lowers the overall flux we expect to receive. As $K_2$ has also decreased, the orbital separation must have also decreased to keep the period constant. A smaller separation and smaller companion mass would suggest the companion's Roche lobe become smaller. The filling factor must then reflect the size of the companion; to find both a lower filling factor and $K_2$ compared to the pre-irradiation models the companion must decrease in size. The nightside temperature remains similar for both approaches, so the lower flux expected from a smaller star on the nightside is compensated for by finding a lower distance.

The filling factor and $K_2$ (through the derived mass ratio) both affect the ellipsoidal component of the companion's optical variability. For example, a larger filling factor produces a more ellipsoidal star, adding flux at the orbital quadrature points ($\phi$ = 0.25, 0.75). If the post-irradiation gravity darkening is moving flux from the centre to the sides of the companion, this in effect removes flux from the superior conjunction whilst adding it to the quadrature points, mimicking ellipsoidal modulation. This relieves the need for a large filling factor to reproduce the observed ellipsoidal component.

The irradiation efficiency, $\epsilon$, is also higher in the post-irradiation model, which is not surprising as heat is more effectively redistributed to the sides but the front of the star still needs to achieve the same temperature in order to reproduce the colours and amplitude at superior conjunction of the companion. For an irradiation-driven redback the irradiation component in the light curve must overcome the comparatively large ellipsoidal component, thus obtaining a high efficiency is not too surprising. Higher efficiencies have only previously been determined for PSR~J1810+1744, an extremely irradiated black widow \citep{Breton2012}. Our pre-irradiation gravity darkening $\epsilon$ is comparable to that found for PSR~J1555$-$2908 \citep{Kennedy2022}. However, much past \texttt{Icarus} modelling has assumed a convective gravity darkening coefficent (0.08) which fundamentally affects the temperature on the companion's irradiated face. The stronger gravity darkening produced by the radiative coefficient deployed here requires more irradiation to achieve the same dayside temperature. In short, irradiation efficiencies of models with varying gravity darkening coefficients should not be directly compared. Post-irradiation gravity darkening then exacerbates this further, as the irradiation itself is gravity darkened. Yet more irradiating flux is then required to reproduce the temperature pattern. This quite naturally accounts for the increased T$_\text{irr}$ and $\epsilon$ for the post gravity darkening models. 

Our modelling does not decisively indicate whether pre- or post-irradiation gravity darkening is preferred. Comparing our D+C models the Bayesian evidence as provided by the \texttt{dynesty} sampler is higher for the pre-irradiation gravity darkening case. The photometric fit is also better. However, post-irradiation gravity darkening models find a much tighter fit to the radial velocity curve. We tentatively support the post-irradiation gravity darkening case over the pre-irradiation gravity darkening due to the improved radial velocity fit in addition to our work as well as similar conclusions obtained by other authors \citep[see][]{Romani2021}. This is also driven from the fact that it probably replicates the physical conditions on the companion's surface, though full scale simulations of an irradiated atmosphere would be required to settle this. In conclusion, we suggest that our post-irradiation gravity darkening D+C model is our `best-fit model' to characterise the companion in this system.

\subsection{Centre-of-light corrections}
\label{s:CoL}
\begin{table}
\centering
\caption{Centre-of-light corrections implied by pre- and post- irradiation gravity darkening D+C models.}
\label{t:CoL}
\begin{tabular}{ll}
\hline Gravity Darkening &  Centre-of-light correction \\ \hline
                         &                \\
Pre-irradiation  &  1.05 $\pm$ 0.06             \\
Post-irradiation  &   1.00 $\pm$ 0.07             \\
                 &                     \\ \hline
\end{tabular}
\end{table}
\label{CoL}
As described in \S\ref{Intro}, surface heating of the companion is expected to affect where a given spectral line is emitted. Thus a centre-of-light correction is needed to get the radial velocity determined for that line to reflect the true centre-of-mass radial velocity, 
\begin{equation}
    k = \frac{K_\text{CoL}}{K_\text{CoM}}
\end{equation}. 

Depending on where exactly the line is emitted, we should expect either a larger or smaller centre-of-light radial velocity than that the centre-of-mass radial velocity; larger if the line is preferentially emitted towards the nightside of the star (effectively orbiting at a larger radius than the CoM), or smaller if the line is stronger on the irradiated dayside. \citet{Linares2018} (hereafter L18) models PSR~J2215+5135, as in this work, using Balmer dominated and Mg$\beta$ radial velocity curves. They calculate the expected equivalent width (EW) of each line across the companion's surface. They conclude the lower temperature Mg$\beta$ line tracks the nightside and the high temperature Balmer series the dayside, `bracketing' the $K_\text{CoM}$ between them. 

Appendix A of \citet{K&R2020} adds some nuance to the `bracketing' scenario. They assert that, whilst the EW of the Mg$\beta$ triplet is indeed highest across the nightside, the raw EW is not the correct metric to use to measure the brightness of a given line. Rather, the EW must be weighted by the continuum flux at that point. A stronger line is not necessarily brighter, the local brightness dominates over the varying line strength over the surface. When weighting the EW by the local flux, the Mg$\beta$ triplet is expected to be brightest towards the dayside, rejecting the `bracketing' scenario. 

Figure \ref{fig:specRV} lends credence to the conclusion of \citet{K&R2020}. The amplitude of our modelled radial velocity curve supports the Mg$\beta$ feature being stronger towards the dayside, or at least does not support observing it toward the nightside, given it has a lower amplitude than the centre-of-mass velocity sampled to generate it. Table \ref{CoL} displays the correction needed for the observed (red) curve. For both gravity darkening prescriptions, the correction is within 1$\sigma$ of coincidence with the centre of mass. The exact value determined is clearly affected by the prescription chosen. Here we can appeal to our physical model. As in L18, we have calculated the EW of the H$\beta$ and Mg$\beta$ triplet across the companion's surface. To standardize our calculation we follow the procedure of \citet{Trager1998}. Here, the flux weighted EW (wEW) is calculated as 
\begin{equation}
    \text{wEW} = {F_C} \int_{\lambda_l}^{\lambda_h} 1 - \frac{F_\lambda}{F_C}\,d\lambda 
    \label{eq:wEW}
\end{equation} relative to a continuum level calculated either side of the spectral feature within predetermined wavelength ranges, and weighted by the continuum level. The wEW for a given line can then be determined for every Icarus surface element, producing a EW map of the surface. 

Figure \ref{fig:surface} shows several absorption line surface maps produced for our D+C models, most notably the temperature and wEW. The temperature maps immediately reinforce differing heating patterns between the two options: applying gravity darkening after irradiation effectively removes flux from the center of the dayside, whilst adding it to the sides of the companion as compared to the pre-irradiation gravity darkening case. The effect this has on the centre-of-light correction is then somewhat predictable. The broader irradiation of the post-irradiation model naturally lowers the correction needed, meaning the Mg$\beta$ triplet more closely tracks the centre-of-mass. Conversely, the sharply heated dayside for the pre-irradiation gravity darkening case concentrates the line flux towards the companion's nose, exacerbating the correction needed. 

Naturally the two line species can also be compared. For H$\beta$ the wEW is clearly higher towards the dayside. The Mg$\beta$ triplet is slightly stronger on the dayside, but relative to H$\beta$ sees a fairly uniform distribution across the surface at all phases. This nicely reflects the expected interplay between the EW and continuum flux; For Mg$\beta$ between the two distributions the whole surface is covered. By weighting the surface element velocities by their wEW we can make an estimate of the correction needed between the centre-of-mass and centre-of-light velocities. A physical interpretation of this is shown on the wEW map for each line: the red dashed line shows the effective centre-of-light position of the line relative to the centre of mass. For H$\beta$, matching the concentration of wEW on the dayside, the centre of light moves much closer to the nose of the star. For Mg$\beta$, we find the centre of light is actually nearly coincident with the centre of mass. Full surface plots including the EW and continuum flux maps are available in appendix \ref{fig:appendixsurface}.

\section{Conclusions}
\label{Concs}
In this work we have presented the discovery, radio timing and multi-wavelength optical photometry of the redback PSR~J1910$-$5320, as well as updating the radial velocity curve reported in \citet{Au2022+J1910}. These datasets have been modelled using \texttt{Icarus}, providing a new neutron star mass measurement. We have also tested our assumptions about the heating in spider systems, in particular examining whether the surface should be gravity darkened before or after the irradiation is applied to the companion.

Our modelling has constrained a number of system parameters. All our models find an inclination consistent with $\sim 46^\circ$, and similar base temperatures consistent with our expectation from the spectral energy distribution. The remaining parameters vary bimodally, depending on whether gravity darkening is applied before or after irradiation. In particular the filling factor, irradiating temperature (and thus efficiency), companion velocity, distance and component masses change depending on our gravity darkening prescription. For both models a moderate pulsar mass is found, constrained to better than $15\%$ fractional uncertainty at the $68\%$ level.

The novel radial velocity modelling deployed here has also provided evidence that, as advanced in \citet{K&R2020}, the centre-of-light position of absorption species is not solely determined by its activation temperature. We find the metallic, low temperature Mg$\beta$ triplet closely tracks the centre-of-mass velocity, balancing the temperature dependence of the EW and continuum flux. This is currently only verified for J1910, an irradiation driven redback, though our findings should also apply to other systems presenting milder irradiation effects.

The modelling performed here aims to be widely applicable to all spiders where photometry can be supplemented with radial velocity curves. Further spider discovery and follow-up, particularly spectroscopic, is then desirable to provide more reliable measurement, taping on better self-consistency in the way that the centre-of-mass is inferred from spectral lines. Whilst J1910 did not yield a `super-massive' neutron star, which can directly constrain the neutron star EoS, the current work adds to the tally of spider masses and can help understand better the evolution landscape between black widows and redbacks, but also across to other types of neutron star binaries.

\section*{Acknowledgments}
R.P.B. acknowledges support from the European Research Council (ERC) under the European Union's Horizon 2020 research and innovation program (grant agreement No. 715051; Spiders). O. G. D acknowledges the support of a Science and Technology Facilities Council (STFC) stipend (Grant number: 2487578) to permit work as a post graduate researcher. E.C.F. is supported by NASA under award number 80GSFC21M0002. V.S.D. and ULTRACAM are funded by the Science and Technology Facilities Council (grant ST/V000853/1). J.S. acknowledges support by NASA grants 80NSSC22K1583 and 80NSSC23K1350, NSF grant AST-2205550, and the Packard Foundation. C.J.C., E. D. B., P.C.C.F., M.K. and P.V.P. acknowledge continuing valuable support from the Max-Planck Society. K.L.L. and K.Y.A. are supported by the National Science and Technology Council of the Republic of China (Taiwan) through grants 111-2636-M-006-024 and 112-2636-M-006-009. K.L.L. is also a Yushan Young Fellow supported by the Ministry of Education of the Republic of China (Taiwan).

The MeerKAT telescope is operated by the South African Radio Astronomy Observatory, which is a facility of the National Research Foundation, an agency of the Department of Science and Innovation. We thank staff at SARAO for their help with observations and commissioning. TRAPUM observations used the FBFUSE and APSUSE computing clusters for data acquisition, storage and analysis. These clusters were funded, designed and installed by the Max-Planck-Institut-für-Radioastronomie (MPIfR) and the Max-Planck-Gesellschaft. FBFUSE performs beamforming operations in real-time using the \texttt{mosaic} software stack \citep{Chen2021+Beamformer}. Observations made use of the Pulsar Timing User Supplied Equipment (PTUSE) servers at MeerKAT which were funded by the MeerTime Collaboration members ASTRON, AUT, CSIRO, ICRAR-Curtin, MPIfR, INAF, NRAO, Swinburne University of Technology, the University of Oxford, UBC and the University of Manchester.  The system design and integration was led by Swinburne University of Technology and Auckland University of Technology in collaboration with SARAO and supported by the ARC Centre of Excellence for Gravitational Wave Discovery (OzGrav) under grant CE170100004.

The Parkes `Murriyang' radio telescope is part of the Australia Telescope National Facility (\url{https://ror.org/05qajvd42}) which is funded by the Australian Government for operation as a National Facility managed by CSIRO. We acknowledge the Wiradjuri people as the Traditional Owners of the Observatory site.

This work has made use of data from the European Space Agency (ESA) mission {\it Gaia} (\url{https://www.cosmos.esa.int/gaia}), processed by the {\it Gaia} Data Processing and Analysis Consortium (DPAC, \url{https://www.cosmos.esa.int/web/gaia/dpac/consortium}). Funding for the DPAC has been provided by national institutions, in particular the institutions participating in the {\it Gaia} Multilateral Agreement.

Based on observations collected at the European Southern Observatory under ULTRACAM GTO program 0109.D-0678. We are grateful to the astronomy and technical staff support at La Silla Observatory who enabled the remote observation program and observer John A. Paice. 

Based on observations obtained at the Southern Astrophysical Research (SOAR) telescope, which is a joint project of the Minist\'{e}rio da Ci\^{e}ncia, Tecnologia e Inova\c{c}\~{o}es (MCTI/LNA) do Brasil, the US National Science Foundation’s NOIRLab, the University of North Carolina at Chapel Hill (UNC), and Michigan State University (MSU).

This research has made use of "Aladin sky atlas" developed at CDS, Strasbourg Observatory, France \citep{AladinDesktop}. 

\section*{Data Availability}
TRAPUM and ULTRACAM observations are available upon reasonable request to the contact author. SOAR/Goodman observations are available from the NOIRLab Astro Data Archive \url{https://astroarchive.noirlab.edu/}.


\bibliographystyle{mnras}
\bibliography{bib} 

\appendix
\section{Radial velocity fitting}
\label{RVfitappendix}
\begin{figure}
    \includegraphics[width = \linewidth]{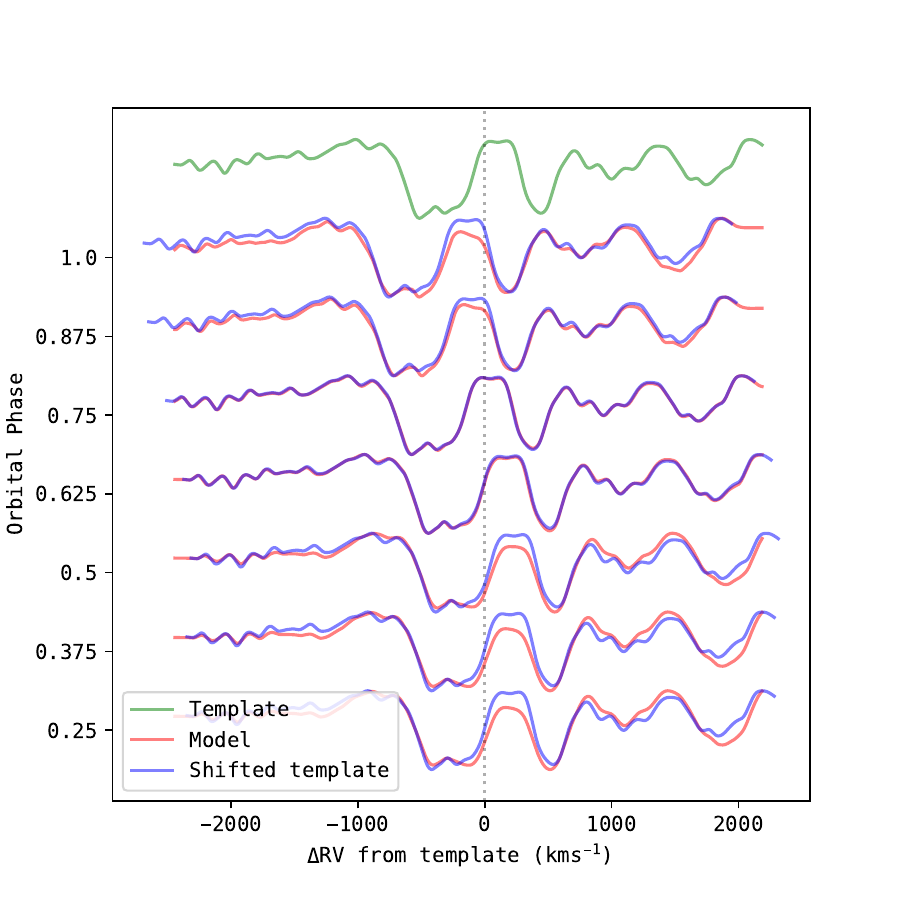}
    \caption{Template (green) Mg$\beta$ spectrum shifted (blue) to determine the effective radial velocity at various orbital phases (red).}
    \label{fig:RVtemplates}
\end{figure}
The radial velocity fitting technique employed here fundamentally aims to take only the most essential information from \texttt{Icarus} spectroscopy modelling. Comparing the model spectra with full observed spectra seems on the surface appealing as the fit can be informed both by the position and depth/profile of a set of lines. Not only is the radial velocity constrained but, in principle, also the temperature. However, systematic effects such as the exact elemental abundances can greatly complicate the situation and drive parameter estimation to compensate by modifying other parameters away from their `true' values. Photometry modelling is not really affected by such considerations as line contribution to the total flux is negligible. Another important challenge to overcome is the considerable computational expense connected to the full modelling of a spectral dataset.

The most essential, model constraining information to extract from a spectrum is the radial velocity, encoded in the Doppler shift of individual lines. This is highlighted particularly in the case of J1910, where we add a likelihood term according to the radial velocity curve rather than the observed spectroscopy directly. Determining radial velocities is, in theory, quite simple: the Doppler shift in a line's wavelength relative to its value at rest reflects the velocity it was emitted at. The wavelength shift should be relatively insensitive to the systematics mentioned above if the overall line shape is not too dissimilar to the template which is being used. For example, we would assume that underpinning our model spectra with atmospheres of differing metallicities should not result in differing radial velocity measurements if we consider one line species at a time. Conversely, the depth of lines would change quite dramatically with metallicity. Thus we can be relatively confident that radial velocities derived from a model can be reliable, even if some of the assumptions regarding abundances are off so long as the temperature profile and stellar and binary parameters are captured adequately (via the photometry), Moreover, as we are only interested in individual lines the computational cost is greatly reduced.

Figure \ref{fig:RVtemplates} demonstrates our simplified spectroscopy modelling and radial velocity fit. Given a radial velocity curve, we generate a synthetic \texttt{Icarus} spectrum for the orbital phases at which radial velocity measurements are available. A reference orbital phase is picked as a template -- either that with the strongest line feature or closest to a user defined phase. This template is then cross-correlated with the others for the wavelength, and thus velocity, shift. This produces a relative radial velocity curve within our model, with the expected sinusoidal shape. We then fit this to the observed curve, analytically minimising a velocity offset, to find the additional likelihood term to the model (via a $\chi^2$ penalty). Even though the radial velocity measurements extracted from the observed spectra in \S \ref{spectroscopy} adopted a standard template profile, our model fitting to the velocity should closely resembles them for the reasons that were explained above.

\section{Supplementary plots}
\begin{figure*}
    \includegraphics[width = \linewidth]{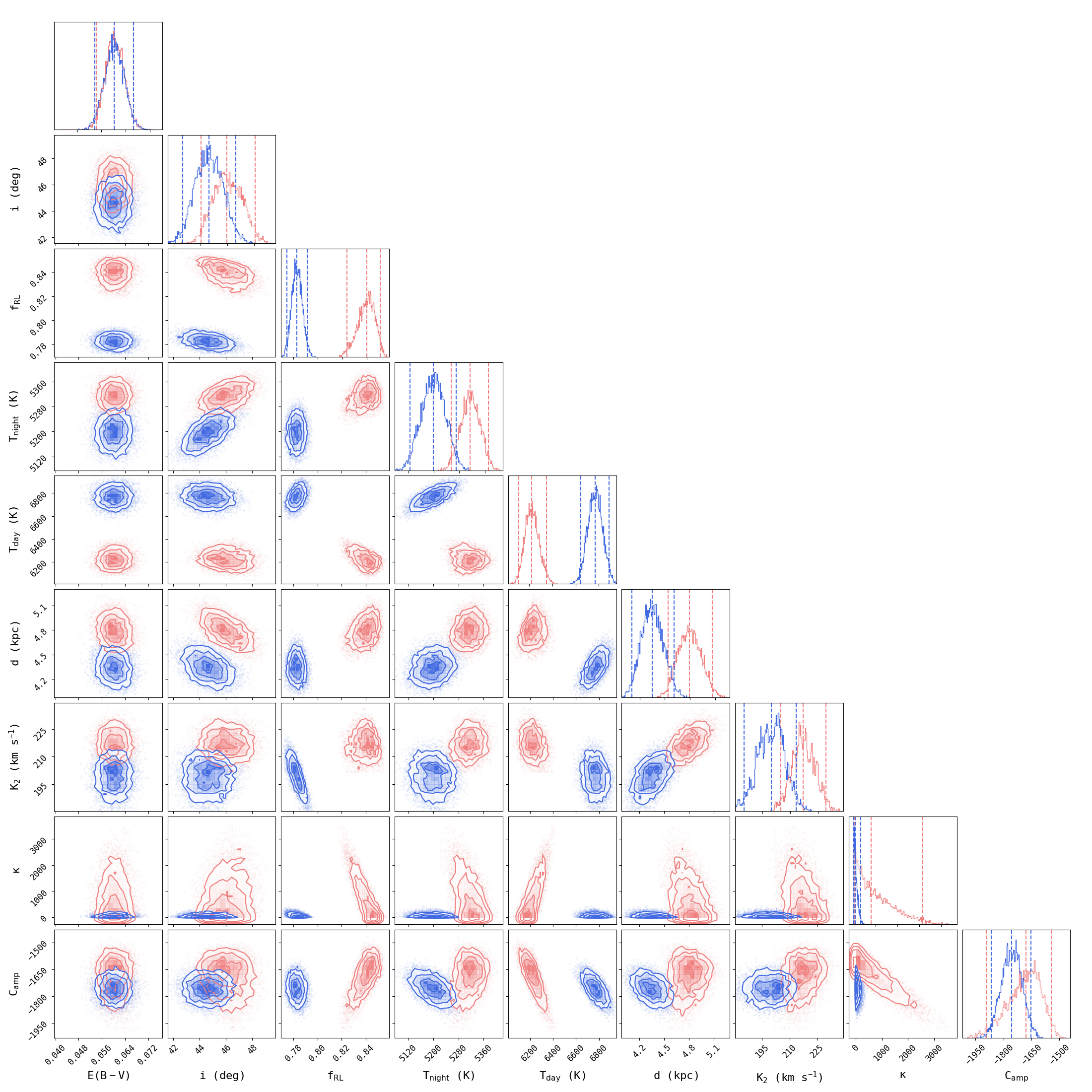}
    \caption{Corner plot showing \texttt{Icarus} fit parameters for pre- (red) and post- (blue) irradiation gravity darkening diffusion + convection models. Contours outline the 68, 95, and 99.7\% confidence intervals. The dashed lines on the 1D posterior plots show the 0.025, 0.5 and 0.975 quantiles.}
    \label{fig:fitcorner}
\end{figure*}

\begin{figure*}
    \includegraphics[width = \linewidth]{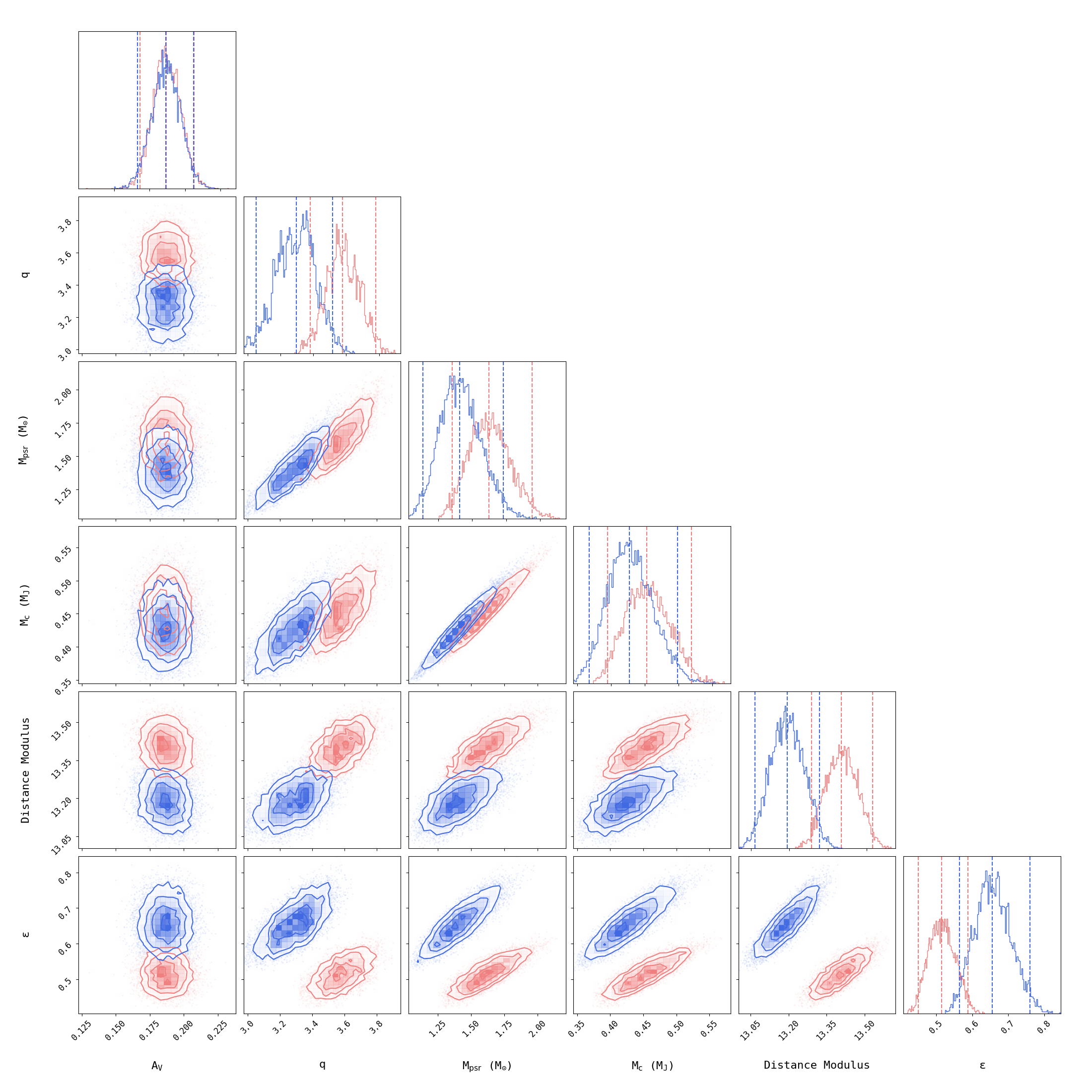}
        \caption{Corner plot showing derived parameters for pre- (red) and post- (blue) irradiation gravity darkening diffusion + convection models. Contours outline the 68, 95, and 99.7\% confidence intervals. The dashed lines on the 1D posterior plots show the 0.025, 0.5 and 0.975 quantiles.}
    \label{fig:derivedcorner}
\end{figure*}

\begin{figure*}
    \includegraphics[width = \linewidth]{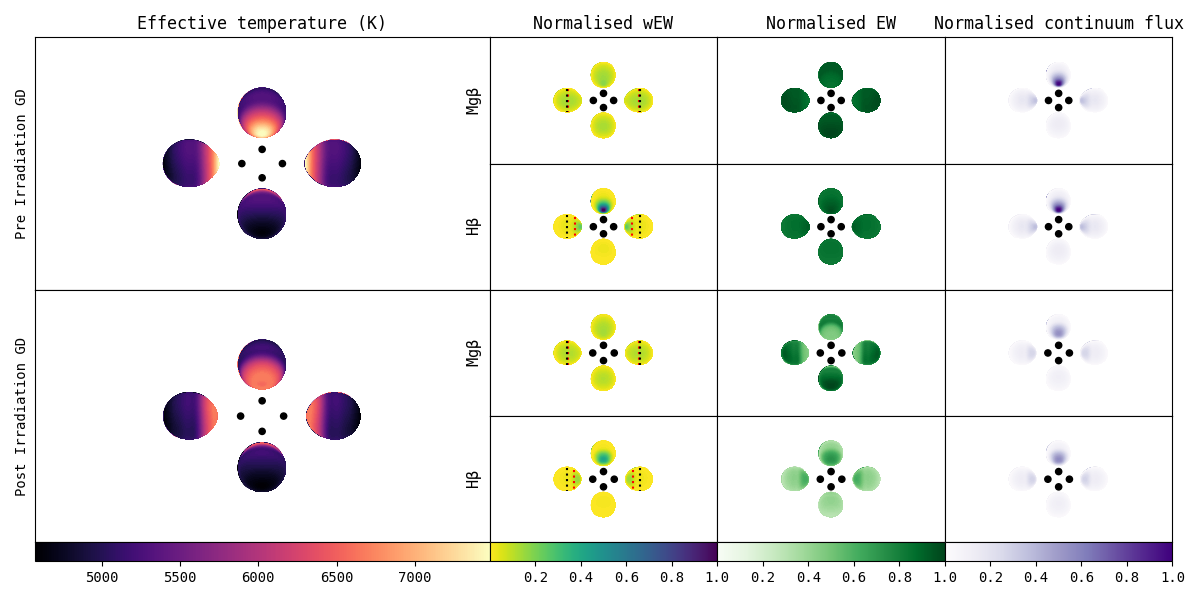}
    \caption{Surface maps for pre- (top) and post- (bottom) irradiation gravity darkening diffusion  + convection models. The leftmost plot shows the surface temperature over the companion surface. The grids on the right show the normalised flux weighted equivalent width (wEW), equivalent width (EW) and continuum flux from each surface element (see (\ref{eq:wEW})). These are split into the Mg$\beta$ triplet, which corresponds with our radial velocity curve, and the H$\beta$ feature. In the picture of \citet{Linares2018} these lines should track the companion nightside and dayside respectively. The dashed lines on the wEW maps indicate the centre of mass (black) and centre of light (red) positions for the given line.}
    \label{fig:appendixsurface}
\end{figure*}
\bsp	
\label{lastpage}
\end{document}